\documentclass[12pt, draftclsnofoot, onecolumn]{IEEEtran}
\ifCLASSINFOpdf
\else
\fi
\usepackage[utf8]{inputenc}
\usepackage{amsmath,amssymb,amsfonts}
\usepackage{cite}

\usepackage{algorithmic}
\usepackage{stfloats}
\usepackage{graphicx}
\usepackage{textcomp}
\usepackage{xcolor}
\usepackage{pifont}
\usepackage{amsthm}
\newtheorem{theorem}{Theorem}

\usepackage{hyperref}
\usepackage{mathrsfs}
\usepackage{booktabs}
\usepackage{hyperref}
\usepackage{cases}
\usepackage{comment}
\usepackage{empheq} 
\usepackage{caption}
\usepackage{subcaption}
\allowdisplaybreaks[4]

\usepackage{url}  %Required
\usepackage{graphicx}  %Required
\usepackage[ruled,vlined,linesnumbered]{algorithm2e}
\usepackage{setspace}
\usepackage{floatpag} 
\usepackage{float}
\floatpagestyle{empty}

\definecolor{red}{RGB}{0,0,0}

\usepackage{listings}

\begin{document}
\captionsetup[figure]{labelfont={rm},labelformat={default},labelsep=period,name={Fig.}}
\title{Joint Localization and Beamforming for Reconfigurable Intelligent Surface Aided 5G mmWave Communication Systems}

\author{Yunis Xanthos,~Wanting Lyu,~Songjie Yang,\\  Chadi Assi,~\IEEEmembership{Fellow,~IEEE},~Xianbing Zou,~Ning Wei,~\IEEEmembership{Member,~IEEE}
% <-this % stops a space
\thanks{Y.~Xanthos),  W.~Lyu, S.~Yang, N.~Wei and X.~Zou are with National Key Laboratory of Science and Technology on Communications, University of Electronic Science and Technology of China, Chengdu 611731, China (E-mail:  
xiuyue12345678@163.com,
lyuwanting@yeah.net,
yangsongjie@std.uestc.edu.cn, zouxb@uestc.edu.cn, wn@uestc.edu.cn).
C. Assi is with Concordia University, Montreal, Quebec, H3G 1M8, Canada (email:assi@mail.concordia.ca).
}

\thanks{This work was supported in part by the National Key R$\&$D Program of China (No. 2020YFB1805000), the National Natural Science Foundation of China (NSFC) under Grant 61871070, 91938202, and the Sichuan Science and Technology Program under Grant 2022YFG0005. The corresponding author is Ning Wei.}}

\maketitle
\begin{abstract}
Reconfigurable intelligent surface  (RIS) is an attractive technology to improve the transmission rate of millimetre-wave (mmWave) communication systems. The previous {research} on RIS technology mainly focused on improving the transmission rate and security rate of the mmWave communication systems. Since the emergence of RIS technology creates the conditions for generating an intelligent radio environment, it also has potential advantages on improving the localization accuracy of the mmWave communication systems. Deployed on walls and objects, RISs are capable of significantly improving communications and positioning coverage by controlling the multi-path reflection. This paper considers the RIS-aided mmWave localization system and proposes a joint beamforming and localization problem. However, since the objective function depends on the unknown UE's position and instantaneous channel state information (CSI), this beamforming and localization technology based on RIS assistance is challenging. To solve this problem, we propose a new joint localization and beamforming optimization (JLBO) algorithm, and give the proof of its convergence. The simulation results show that the RIS can improve the user localization accuracy of the system and the proposed scheme has a significant performance improvement compared with the traditional schemes.
\end{abstract}

\begin{IEEEkeywords}
Reconfigurable intelligent surface, millimeter-wave,  channel states information, localization, beamforming. 
\end{IEEEkeywords}

% \fi
\IEEEpeerreviewmaketitle

\section{Introduction}
\IEEEPARstart{D}{ue} to the increasing demands for localization-based services in fifth-generation mobile communications (5G) and beyond 5G (B5G) systems, especially millimeter-wave (mmWave) and terahertz wireless networks, the localization technology is becoming more and more important. In \cite{8755880}, the authors have demonstrated that the positioning accuracy of mmWave systems can reach centimeter-level accuracy. Studies have shown that information from non-line-of-sight (NLOS) paths can be effectively used in mmWave systems to estimate the {location} of physical scatters or reflectors, enabling simultaneous localization and mapping of radios over time propagation environment \cite{9367007}. Many works have studied the localization of mmWave wireless communication based on beamforming technology. In\cite{6799306}, the author explored the improvement of positioning accuracy of mmWave distributed antenna systems via beamforming technology.  Authors in\cite{6125211} investigated the improvement of mmWave distributed antenna system localization accuracy through actual beamforming.  In\cite{8356190}, the system transmission rate was improved with reduced localization error by optimizing radio frequency resources.  In\cite{8370829}, the UE's position was estimated first by determining the UE's direction and then by beam alignment.  In\cite{8240645}, the authors studied the determination of uplink and downlink UE and base station (BS) {location} of the wireless communication system with a 3D mmWave channel model. However, the above literature mainly focused on realizing UE localization only through active beamforming. Recently, with the emergence of RIS, passive beamforming technology has been widely used to solve some challenges in radio positioning \cite{9497709}.

RIS originates from the idea of manipulating the wireless communication environment \cite{9140329}. RIS is a two-dimensional planar reflection array consisting of many low-cost passive reflecting units that can produce a reconfigurable phase shift on the incident signal and then reflect it to the receiving terminal. Because it can usually be manufactured with cheap positive and negative intrinsic (PIN) diodes or varactor diodes and can be deployed almost anywhere to build powerful virtual line-of-sight (VLOS) links without the need for power-consuming radio frequency (RF) chains \cite{9140329}, it is considered to be a promising hardware solution that can solve the problems of propagation limitations, hardware cost and energy consumption.

\textcolor{red}{There have been many papers studying the RIS-aided localization problem in different scenarios. According to the scenarios, these papers can be divided into two categories, namely near-field \cite{9650561,9133126,9206044} and far-field \cite{9576697,9684431,8732647}. Depending on where the RIS is deployed, it can be further classified as being placed on the BS side \cite{8755880,9053493}, on the UE side \cite{9215972}, or as a separate reflector \cite{9528041,9500281}. According to the number of antennas, it can be divided into multiple input multiple output (MIMO) \cite{9614196,9743440}, multiple input single output (MISO) \cite{9615188}, and single input single output (SISO) \cite{9745078,9438669}, etc. In addition, RIS can be used to assist radar systems in improving target detection capabilities \cite{9348016,9650561}. Authors in \cite{9593204} established the Cram$\mathop{\mathrm{e}}\limits^{\prime}$r-Rao lower bound (CRLB) of continuous RIS near-field localization error and studied the influence of RIS size on the localization error. The effect of limited RIS phase resolution was investigated in \cite{9508872} for the same scenario. Furthermore, the authors considered the localization problem with RIS as a lens and proposed an estimation algorithm \cite{9048973}. In \cite{abu2021near,rahal2021ris}, the authors studied the near-field localization of a BS and a RIS, where it was shown that the system can accurately obtain the UE's {location} information even if the direct path from the BS to the UE is blocked. In \cite{liu2021reconfigurable,elzanaty2021reconfigurable}, considering RIS-aided MIMO scenarios, the authors derived CRLB to optimize the RIS phase distribution for localization. Moreover, it has been demonstrated that positioning and UE synchronization can be performed with a single RIS in MISO \cite{fascista2021ris} and even in SISO \cite{keykhosravi2021siso} settings when far-field conditions remain unchanged. In addition, \cite{nnamani2021joint} and \cite{xing2021location} studied unmanned aerial vehicle (UAV)-RIS-assisted system positioning and beamforming, and the author proposed an alternating optimization algorithm based on Karush–Kuhn–Tucker (KKT) conditions. In addition, the positioning accuracy of CRLB was deduced in \cite{xing2021location}. Authors in \cite{teng2022bayesian} studied user localization and tracking, and proposed a joint location pursuit algorithm based on Bayesian estimation. In \cite{cheng2022ris}, the location of the user was estimated on the assumption that the RIS location is known. In  \cite{gao2022joint} and \cite{hu2022irs}, the joint localization and beamforming design in simultaneously transmitting (STAR)-RIS-assisted non-orthogonal multiple access (NOMA) system was investigated. Authors in \cite{guo2022joint} considered the estimation of user location with the goal of maximizing the security rate.
Summing up the above papers, we find that although the existing works have studied the localization algorithms of RIS-assisted mmWave systems, most of the papers assumed that the location of RIS is known and only considered user location estimation and there is no further research on beamforming design when the RIS location information is known. In addition, it is observed that although some papers have given the CRLB of the localization accuracy to evaluate the performance, none of them exploited the CRLB as the optimization goal, resulting in uncontrollable estimation accuracy and CRLB gap.}

Therefore, to fill these gaps, we study the location and beamforming of RIS-aided millimeter wave systems without RIS position information, and we propose a BCD algorithm based on CRLB optimization and SCA method. Specifically, in this paper, we mainly research beamforming technology and RIS technology to improve localization accuracy. However, the positioning problem in RIS-assisted mmWave communication systems is challenging. First, in the objective function based on the positioning error, the value of the error function is determined by the beamforming vector, the RIS reflection matrix, the CSI and the UE position. However, the UE {location} and CSI are often unknown in the existing system. {\color{red}Second}, because the localization error function is a function of RIS reflection matrix and beamforming, and the unit modulus limitation of RIS reflection matrix and multivariable coupling make the problem non-convex, the traditional brute force testing algorithm needs a very high complexity to deal with this kind of problem.

In \cite{praia2021phase}, power and bandwidth optimization for localization (by optimizing the CRLB) was studied, where the UE {location} is assumed to be known. To overcome the first challenge, a solution from a robust optimization method was used to optimize the radio resource for localization \cite{keykhosravi2022ris,9593167}. Specifically, a worst-case CRLB with respect to (w.r.t.) an uncertainty set of UE {location} parameters is used as the optimization objective. However, such a robust optimization method is usually over-conservative, especially when the uncertainty set is large. This uncertainty jeopardizes the associated performance gain. To overcome the non-convexity challenge, successive convex approximation (SCA) or majorization minimization (MM) methods are commonly used \cite{8502790,8752072,8736776}. However, the brute-force application of these algorithms will result in poor performance.

This paper mainly studies the problem of RIS-aided mmWave localization and beamforming. To solve this problem, we propose a joint localization and beamforming optimization (JLBO) algorithm based on the block coordinate descent (BCD) method, which does not need to know the {\color{red} location} of the UE. Specifically, the contributions of this paper are summarized as follows:
\begin{itemize}
\item
In addition to the traditional localization system, this paper uses RIS technology. Since RIS introduces extra degrees of freedom, {\color{red} not only can it improve the estimation accuracy} of position parameters but also achieve beam alignment between the BSs and UE. In addition, because we take CRLB as the optimization goal in the beamforming design phase, the localization accuracy of the RIS-aided mmWave system is further improved.
\item
In this paper, our proposed JLBO algorithm is an alternating optimization algorithm based on three sub-problems: channel gain estimation, beamforming design and {location} parameter estimation. The algorithm does not need to know the RIS's {location} or the UE's prior spatial information. Because we solve each optimization problem according to the optimal algorithm, compared with the traditional triangulation localization method based on TOA and AOA estimation, the proposed position parameter estimation algorithm can not only estimate the position of the UE and the scatterer but also improve the performance of the system.
\item
In this paper, to solve the non-convex problem, we propose a new beamforming design algorithm based on non-convex approximation method. Since the problem structure in this paper can not directly use the conventional SCA algorithm, we propose a non-convex successive approximation algorithm based on the SCA algorithm. Finally, we prove the convergence of the algorithm, and the numerical results show that it performs better than the existing localization algorithms.
\end{itemize}

\textbf{Organization:} This paper contains the following sections. In section \ref{II}, the system model and problem formulation are introduced. In section \ref{III}, the JLBO algorithm is presented. Specifically, this section includes channel gain estimation, beamforming design, and position parameter estimation. Section \ref{IV} analyzes the convergence of the proposed algorithm  Section \ref{V} proves the performance of the proposed algorithm by using numerical results  Section \ref{VI} summarizes the paper.

\section{System Model and Problem Formulation}\label{II}
\begin{figure}
\begin{subfigure}{.5\textwidth}
  \centering
  % include first image
  \includegraphics[scale=0.35]{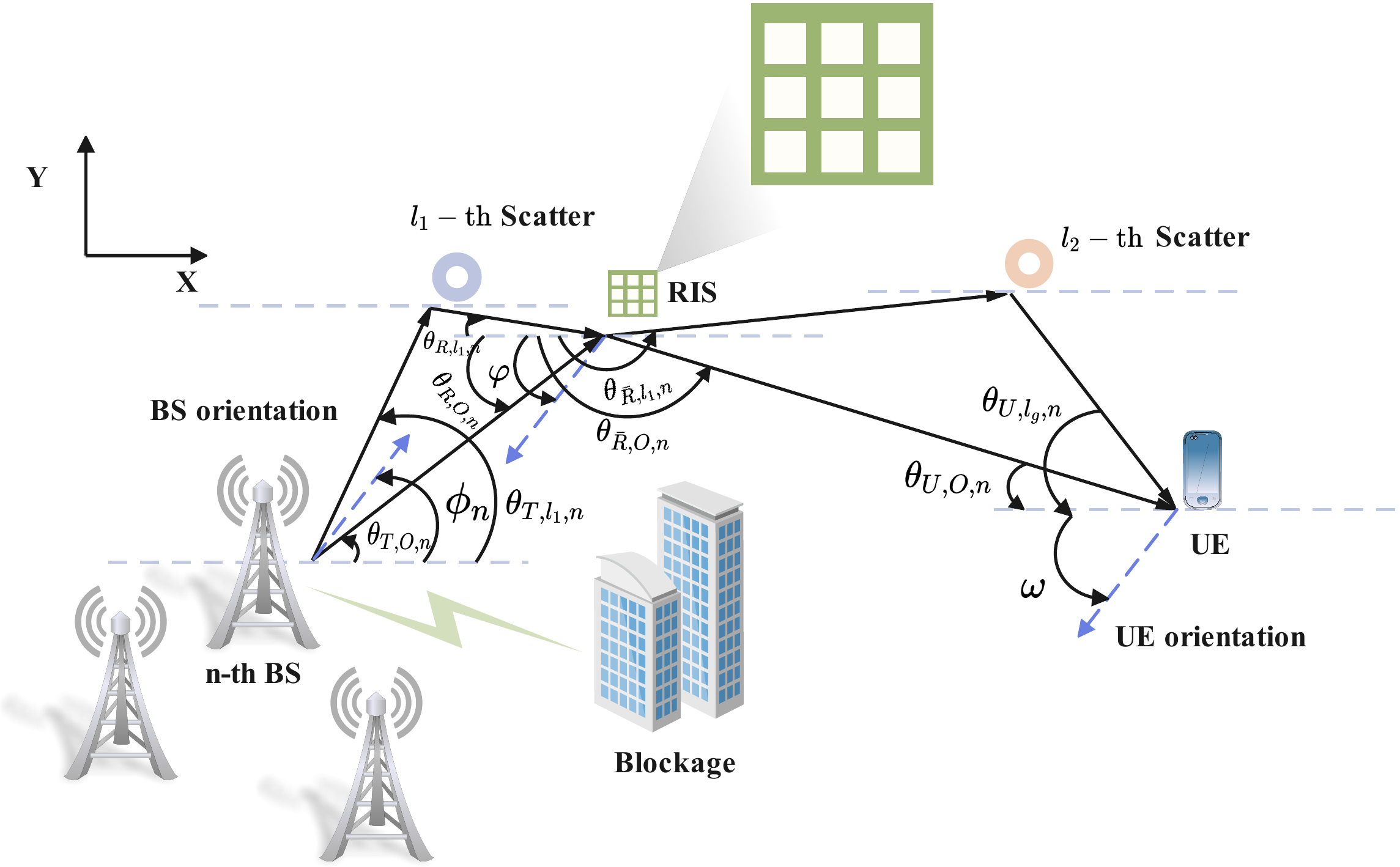}  
  \caption{Illustration of the RIS-aided mmWave wireless localization system.}
  \label{fig:(3-10a)}
\end{subfigure}
\begin{subfigure}{.5\textwidth}
\centering
\includegraphics[scale=0.35]{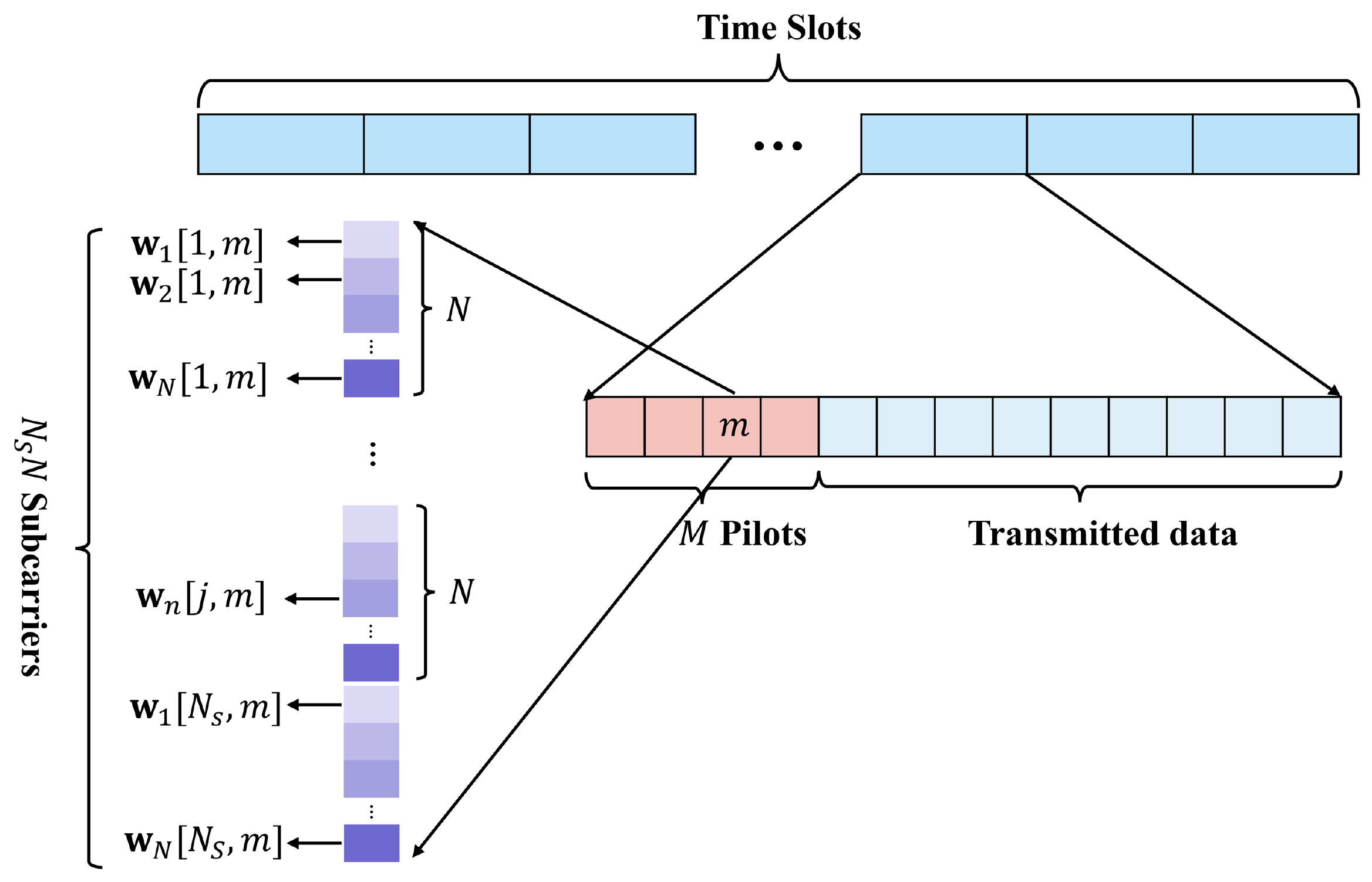}
\caption{Pilot signal structure and the subcarrier {\color{red}allocation}.}
\label{fig:3_11}
\end{subfigure}
\caption{ Illustration of the system model and pilot signal structure.}
\label{system_model_1}
\end{figure}

As shown in Fig.\ref{system_model_1}(a), a RIS-aided mmWave multiple access points (BSs) communication system is considered, where the system includes $N$ BSs and one UE. The number of the antennas of each BS and the number of the antennas of the UE are $N_{T}$ and $N_{U}$, respectively. In addition, the BSs and the UE are equipped with a uniform linear array (ULA), respectively. To obtain the channel state information (CSI) and the {location} of the UE, we give the downlink pilot signal in Fig.\ref{system_model_1}(b). Since the pilot signal structure is designed based on CSI, in the following, we will describe the mmWave channel model and pilot signal structure in detail. 

\subsection{MmWave Channel Model and UE Location Model}
In this paper, the mmWave channel model-based limited scattering model is considered. The mmWave channel from the $n$-th BS to the RIS is expressed as 
\begin{align}
\mathbf{G}_{n}[t,j]=\mathbf{A}_{R,n}[j]\tilde{\mathbf{G}}_{n}[t,j]\mathbf{A}_{T,n}^{H}[j],\label{formu1}
\end{align}
where $\mathbf{G}_{n}[t,j]\in\mathbb{C}^{N_{R}\times N_{T}}$, $j$ and $t$ denote the $j$-th subcarrier and the $t$-th slot, respectively. The mmWave channel from the RIS to the UE is expressed as
\begin{align}
\mathbf{H}_{n}[t,j]=\mathbf{A}_{U,n}[j]\tilde{\mathbf{H}}_{n}[t,j]\mathbf{A}_{\bar{R},n}^{H}[j],\label{formu2}
\end{align}
where $\mathbf{H}_{n}[t,j]\in\mathbb{C}^{N_{U}\times N_{R}}$, $N_{R}$, $N_{U}$ and $N_{T}$ denote the number of RIS reflection elements, UE antennas and BS antennas, respectively. $\tilde{\mathbf{G}}_{n}[t,j]\in\mathbb{C}^{(L_{1}+1)\times (L_{1}+1)}$ and $\tilde{\mathbf{H}}_{n}[t,j]\in\mathbb{C}^{(L_{2}+1)\times (L_{2}+1)}$
are the frequency-domain channel matrices on the $j$-th subcarrier. {\color{red} $L_{1}$ and $L_{2}$ denote the number of channel paths} $\mathbf{A}_{R,n}[j]\in\mathbb{C}^{N_{R}\times (L_{1}+1)}$, $\mathbf{A}_{T,n}[j]\in\mathbb{C}^{N_{T}\times(L_{1}+1)}$ , $\mathbf{A}_{U,n}[j]\in\mathbb{C}^{N_{U}\times (L_{2}+1)}$ and $\mathbf{A}_{\bar{R},n}[j]\in\mathbb{C}^{N_{R}\times (L_{2}+1)}$ denote the steering matrices of the BS, the RIS, and the UE, respectively, which are expressed as
\begin{align}
&\mathbf{A}_{T,n}[j]=[\mathbf{a}_{T,j}(\theta_{T,0,n}),\ldots,\mathbf{a}_{T,j}(\theta_{T,L_{1},n})], \mathbf{a}_{T,j}(\theta_{T,l_{1},n})\in\mathbb{C}^{N_{T}\times 1}=\left[1,\ldots,e^{-\bar{j}\frac{d_{T}\pi}{\lambda_{j}}(N_{T}-1)\sin(\theta_{T,l_{1},n})}\right]^{T},\nonumber\\
&\forall ~l_{1}\in\{0,\ldots,L_{1}\},\forall ~n\in\{1,\ldots,N\},\forall ~j\in\{n,n+N,\ldots,n+(N_{S}-1)N\},\nonumber
\\
&\mathbf{A}_{R,n}[j]=[\mathbf{a}_{R,j}(\theta_{R,0,n}),\ldots,\mathbf{a}_{R,j}(\theta_{R,L_{1},n})], \mathbf{a}_{R,j}(\theta_{R,l_{1},n})\in\mathbb{C}^{N_{R}\times 1}=\left[1,\ldots,e^{-\bar{j}\frac{d_{R}\pi}{\lambda_{j}}(N_{R}-1)\sin(\theta_{R,l_{1},n})}\right]^{T},\nonumber\\
&\forall ~l_{1}\in\{0,\ldots,L_{1}\},\forall ~n\in\{1,\ldots,N\},\forall ~j\in\{n,n+N,\ldots,n+(N_{S}-1)N\},\nonumber\\
&\mathbf{A}_{U,n}[j]=[\mathbf{a}_{U,j}(\theta_{U,0,n}),\ldots,\mathbf{a}_{U,j}(\theta_{U,L_{2},n})], \mathbf{a}_{U,j}(\theta_{U,l_{2},n})\in\mathbb{C}^{N_{U}\times 1}=\left[1,\ldots,e^{-\bar{j}\frac{d_{U}\pi}{\lambda_{j}}(N_{U}-1)\sin(\theta_{U,l_{2},n})}\right]^{T},\nonumber\\
&\forall ~l_{2}\in\{0,\ldots,L_{2}\},\forall ~n\in\{1,\ldots,N\},\forall ~j\in\{n,n+N,\ldots,n+(N_{S}-1)N\},\nonumber\\
&\mathbf{A}_{\bar{R},n}[j]=[\mathbf{a}_{\bar{R},j}(\theta_{\bar{R},0,n}),\ldots,\mathbf{a}_{\bar{R},j}(\theta_{\bar{R},L_{2},n})], \mathbf{a}_{\bar{R},j}(\theta_{\bar{R},l_{2},n})\in\mathbb{C}^{N_{R}\times 1}=\left[1,\ldots,e^{-\bar{j}\frac{d_{R}\pi}{\lambda_{j}}(N_{R}-1)\sin(\theta_{\bar{R},l_{2},n})}\right]^{T},\nonumber\\
&\forall ~l_{2}\in\{0,\ldots,L_{2}\},\forall ~n\in\{1,\ldots,N\},\forall ~j\in\{n,n+N,\ldots,n+(N_{S}-1)N\},\label{formu6}%
\end{align}
where $\bar{j}=\sqrt{-1}$, parameters $\theta_{T,l_{1},n}$, $\theta_{R,l_{1},n}$,  $\theta_{U,l_{2},n}$ and $\theta_{\bar{R},l_{2},n}$ denote the angle of departure (AoD) of the BS, the angle of arrival (AoA) of the RIS, the AoD of the RIS, and the AoA of the UE, respectively. $\lambda_{j}$ denotes the wavelength of the $j$-th subcarrier. $d_{B}$, $d_{R}$ and $d_{U}$ are known scalars denoting the distances between the antenna elements of each BS, each UE and each RIS reflection element, respectively. {\color{red}$\tilde{\mathbf{G}}_{n}[t,j]$} and {\color{red}$\tilde{\mathbf{H}}_{n}[t,j]$}are expressed as
\begin{align}
\tilde{\mathbf{G}}_{n}[t,j]=\sqrt{N_{T}N_{R}}\mathrm{diag}\{g_{l_{1},n}[t]e^{-\bar{j}2\pi\frac{j}{N_{S}T_{S}}\tau_{l_{1},n}}|\forall~l_{1}=0,\ldots,L_{1}\},\nonumber\\
\tilde{\mathbf{H}}_{n}[t,j]=\sqrt{N_{R}N_{U}}\mathrm{diag}\{h_{l_{2},n}[t]e^{-\bar{j}2\pi\frac{j}{N_{S}T_{S}}\tau_{l_{2},n}}|\forall~l_{2}=0,\ldots,L_{2}\},\label{formu8}
\end{align}
where $\bar{j}=\sqrt{-1}$, $g_{l_{1},n}[t]$ and $h_{l_{2},n}[t]$ denote the complex channel gain of the $l_{1}$-th path and $l_{2}$-th path, and the sampling period is denoted as $T_{S}$. $\tau_{l_{1},n}$ and $\tau_{l_{2},n}$ represent the time of arrival (ToA) of the BS to the RIS and the ToA of the RIS to the UE. $N_{S}$ is the number of subcarriers. According to \cite{8624270}, $g_{l_{1},n}[t]\sim\mathcal{CN}(0,\sigma_{l_{1},n}^{2})$ and $h_{l_{2},n}[t]\sim\mathcal{CN}(0,\sigma_{l_{2},n}^{2})$ are independent and identically distributed over different slots, where $\sigma_{l_{1},n}^{2}$ and $\sigma_{l_{2},n}^{2}$ denote the variances of the mmWave channels, which are usually known in practical systems\cite{8624270}.

It is not difficult to find that among the parameters of the channel model, the set of the {location} parameters that determine the {\color{red} location} of the UE are $\boldsymbol{\kappa}\in\mathbb{R}^{2N(L_{1}+1)+2N(L_{2}+1)+6}=[\mathbf{r}^{T},\mathbf{u}^{T},\varphi,\omega,\mathbf{b}^{T},\mathbf{x}^{T}]^{T}$, where $\mathbf{r}\in\mathbb{R}^{2N(L_{1}+1)\times 1}=[\mathbf{r}_{1,1}^{T},\ldots,\mathbf{r}_{1,N}^{T},\ldots,\mathbf{r}_{L_{1}+1,1}^{T},\ldots,\mathbf{r}_{L_{1}+1,N}^{T}]^{T}$ and $\mathbf{u}\in\mathbb{R}^{2N(L_{2}+1)\times 1}=[\mathbf{u}_{1,1}^{T},\ldots,\mathbf{u}_{1,N}^{T},\ldots,\mathbf{u}_{L_{2}+1,1}^{T},\ldots,\mathbf{u}_{L_{2}+1,N}^{T}]^{T}$ are the set of unknown scatter {location} parameters. $\mathbf{b}\in\mathbb{R}^{2\times 1}=[b_{1},b_{2}]^{T}$ and $\mathbf{x}\in\mathbb{R}^{2\times 1}=[x_{1},x_{2}]^{T}$ denote the {location} parameters of the RIS and the UE. Let $\mathbf{e}_{X}\in\mathbb{R}^{2\times 1}=[1,0]^{T}$ and $c$ denotes the light speed. According to Cartesian geometry, these parameters $\{\tau_{l_{1},n},\tau_{l_{2},n},\theta_{T,l_{1},n},\theta_{R,l_{1},n},\theta_{U,l_{2},n},\theta_{\bar{R},l_{2},n}\}$ can be expressed as
\begin{align}
&\tau_{l_{1},n}=\left\{
 \begin{array}{lr}\frac{\|\mathbf{b}-\mathbf{a}_{n}\|^{2}}{c}, & l_{1}=0,\\
\frac{\|\mathbf{b}-\mathbf{r}_{l_{1},n}\|^{2}+\|\mathbf{r}_{l_{1},n}-\mathbf{a}_{n}\|^{2}}{c}, & l_{1}>0,  
\end{array}
\right.
\tau_{l_{2},n}=\left\{
 \begin{array}{lr}\frac{\|\mathbf{x}-\mathbf{b}\|^{2}}{c}, & l_{2}=0,\\
\frac{\|\mathbf{x}-\mathbf{u}_{l_{2},n}\|^{2}+\|\mathbf{u}_{l_{2},n}-\mathbf{b}\|^{2}}{c}, & l_{2}>0,  
\end{array}
\right.\nonumber\\
&\theta_{T,l_{1},n}=
\left\{
 \begin{array}{lr}\arccos\left(\frac{(\mathbf{b}-\mathbf{a}_{n})^{T}\mathbf{e}_{X}}{\|\mathbf{b}-\mathbf{a}_{n}\|^{2}}\right)-\phi_{n}, & l_{1}=0,\\
\arccos\left(\frac{(\mathbf{r}_{l_{1},n}-\mathbf{a}_{n})^{T}\mathbf{e}_{X}}{\|\mathbf{r}_{l_{1},n}-\mathbf{a}_{n}\|^{2}}\right)-\phi_{n}, & l_{1}>0,  
\end{array}
\right.\nonumber\\
&\theta_{U,l_{2},n}=
\left\{
 \begin{array}{lr}\arccos\left(\frac{(\mathbf{x}-\mathbf{b})^{T}\mathbf{e}_{X}}{\|\mathbf{x}-\mathbf{b}\|^{2}}\right)-\varphi, & l_{2}=0,\\
\arccos\left(\frac{(\mathbf{u}_{l_{2},n}-\mathbf{b})^{T}\mathbf{e}_{X}}{\|\mathbf{u}_{l_{2},n}-\mathbf{b}\|^{2}}\right)-\varphi, & l_{2}>0,
\end{array}
\right.\nonumber\\
&\theta_{R,l_{1},n}=\left\{
 \begin{array}{lr}\pi+\arccos\left(\frac{(\mathbf{b}-\mathbf{a}_{n})^{T}\mathbf{e}_{X}}{\|\mathbf{b}-\mathbf{a}_{n}\|^{2}}\right)-\varphi, & l_{1}=0,\\
\pi+\arccos\left(\frac{(\mathbf{b}-\mathbf{r}_{l_{1},n})^{T}\mathbf{e}_{X}}{\|\mathbf{b}-\mathbf{r}_{l_{1},n}\|^{2}}\right)-\varphi, & l_{1}>0,  
\end{array}
\right.\nonumber\\
&\theta_{\bar{R},l_{2},n}=\left\{
 \begin{array}{lr}\pi+\arccos\left(\frac{(\mathbf{x}-\mathbf{b})^{T}\mathbf{e}_{X}}{\|\mathbf{x}-\mathbf{b}\|^{2}}\right)-\omega, & l_{2}=0,\\
\pi+\arccos\left(\frac{(\mathbf{x}-\mathbf{u}_{l_{2},n})^{T}\mathbf{e}_{X}}{\|\mathbf{x}-\mathbf{u}_{l_{2},n}\|^{2}}\right)-\omega, & l_{2}>0,  \label{formu11}
\end{array}
\right.
\end{align}
where $\mathbf{a}_{n}\in\mathbb{R}^{2\times 1}$ is the {location} parameters of the $n$-th BS and it is known by the UE. $\phi_{n}$, $\varphi$ and $\omega$ denote the BS orientation of the $n$-th BS, the RIS orientation of the RIS and the UE orientation of the UE, respectively.

\subsection{Pilot Signal Structure}
{\color{red} The structure of the pilot signal is shown} in Fig.\ref{system_model_1} (b). In this paper, we consider two quasi-static channels, that is, the complex channel gain, namely ToA, AoD and AoA of the millimeter wave channel are assumed to be constant during the channel coherence time. 
Generally, the coherence time of quasi-static channel consists of several slots, where each slot consists of a large number of symbols. The first $M$ symbols are used to train transmit beamforming vectors to jointly estimate channel and UE positions, and the rest of the symbols are used to transmit data. 
In Fig.\ref{system_model_1} (b), $\mathbf{w}_{n}[j,m]\in\mathbb{C}^{N_{T}\times 1}$ represents the $m$-th beamforming vector of the $n$-th BS on the $j$-th subcarrier.

In this paper, we consider that each UE is equipped with the frequency division coordinate multipoint transmission technology in \cite{6146494}, which will ensure the efficiency of the proposed algorithm. Specifically, we use a predefined scheduling scheme to allocate $N_{\bar{S}}$ subcarriers to $N$ BSs {\color{red}equally}, and the number of subcarriers per BS is $N_{S}=\frac{N_{\bar{S}}}{N}$.
Therefore, each BS uses $N_{S}$ subcarriers to obtain the beamforming design. In this paper, we assume that the subcarrier sequence of the $n$-th BS is denoted as
\begin{align}
\{n,n+N,\ldots,n+(N_{S}-1)N\}  \label{formu12}  
\end{align}
The set of the training beamforming vectors is denoted as $\mathbf{w}\in\mathbb{C}^{N_{T}NN_{S}M\times 1}=\mathrm{vec}[\mathbf{w}_{n}[j,m]|\forall n\in\{n,n+N,\ldots,n+(N_{S}-1)N\}, \forall m=1,\ldots,M, \forall n=1,\ldots,N]$. It should be noted that the realization of beamforming vector $\mathbf{w}$ will be determined at each beamforming training stage, and thus it will be viewed as an unknown parameter to be optimized in the beamforming training stage. Once it is optimized, it will keep invariant at the channel estimation stage. 

\subsection{Signal Model}
The $j$-th subcarrier received signal from the $n$-th BS at the $t$-th slot can be expressed as
\begin{align}
\mathbf{y}_{n}[t,j,m]=\mathbf{H}_{n}[t,j]\mathbf{\Theta}\mathbf{G}_{n}[t,j]\mathbf{w}_{n}[j,m]+\mathbf{n}_{n}[t,j,m],\label{formu13}
\end{align}
where $\mathbf{y}_{n}[t,j,m]\in\mathbb{C}^{N_{U}\times 1}$ and $\mathbf{n}_{n}[t,j,m]\in\mathbb{C}^{N_{U}\times 1}$ denote the Additive White Gaussian Noise (AWGN) at the UE. In general, we assume $\mathbf{n}_{n}[t,j,m]\sim\mathcal{CN}(\mathbf{0},\sigma_{n}^{2}\mathbf{I})$, where $\sigma_{n}^{2}$ is the noise variance. $\boldsymbol{\Theta}=\mathrm{diag}\{\boldsymbol{\theta}\}\in\mathbb{C}^{N_{R}\times N_{R}}$, $\boldsymbol{\theta}=[e^{j\theta_{1}},\ldots,e^{j\theta_{N_{R}}}]^{T}\in\mathbb{C}^{N_{R}\times 1}$ is the passive beamforming matrix of the RIS. Since the {location} parameters are unknown and they included in channel matrix $\mathbf{H}_{n}[t,j]$ and  $\mathbf{G}_{n}[t,j]$, $\mathbf{H}_{n}[t,j]$ and  $\mathbf{G}_{n}[t,j]$ are unknown. In addition, since channel matrix $\mathbf{H}_{n}[t,j]$ and channel matrix $\mathbf{G}_{n}[t,j]$ are cascaded channel model in \cite{9328485}, the received signal can be viewed as a received signal with bi-linear variable. In the later part of the algorithm, because we use the alternating optimization algorithm, we give the received signal expressions given $\mathbf{G}_{n}[t,j]$ and $\mathbf{H}_{n}[t,j]$.
When $\mathbf{G}_{n}[t,j]$ is given, the received signal is rewritten as
\begin{align}
\mathbf{y}[t]=\mathbf{\Gamma}(\boldsymbol{\kappa}_{2},\boldsymbol{\theta},\mathbf{w},\mathbf{g}[t])\mathbf{h}[t]+\mathbf{n}[t],\label{formu14}
\end{align}
{\color{red} where $\boldsymbol{\kappa}_{2}\in\mathbb{R}^{(2N(L_{2}+1)+3)\times 1}=[\mathbf{b}^{T},\omega,\mathbf{u}^{T}]$,}
 \begin{align}
&\mathbf{\Gamma}(\boldsymbol{\kappa}_{2},\boldsymbol{\theta},\mathbf{w},\mathbf{g}[t])\in\mathbb{C}^{N_{U}N_{S}NM\times N(L_{2}+1)}=\left[\begin{matrix}
\boldsymbol{\Gamma}_{1}&\ldots&\mathbf{0}\\
\vdots&\ddots&\vdots\\
\mathbf{0}&\ldots&\boldsymbol{\Gamma}_{N}
  \end{matrix}
  \right],\label{formu_15}\\
&\boldsymbol{\Gamma}_{n}\in\mathbb{C}^{N_{U}N_{S}M\times(L_{2}+1)}=[\boldsymbol{\gamma}_{n}^{(1)}[1,1],\ldots,\boldsymbol{\gamma}_{n}^{(N_{U})}[N_{S},M]]^{T},\nonumber\\
&\boldsymbol{\gamma}_{n}^{(r)}[j,m]\in\mathbb{C}^{(L_{2}+1)\times 1}=[\gamma_{0,n}^{(r)}[j,m],\ldots,\gamma_{L_{2},n}^{(r)}[j,m]]^{T},\gamma_{l_{2},n}^{(r)}[j,m]=\mathbf{b}_{n}^{H}[t,j,m]\boldsymbol{\mu}_{l_{2},j,n}^{(r)},\nonumber\\
&\mathbf{b}_{n}[t,j,m]\in\mathbb{C}^{N_{R}\times 1}=\boldsymbol{\Theta}\mathbf{G}_{n}[t,j]\mathbf{w}_{n}[j,m],\boldsymbol{\mu}_{l_{2},j,n}^{(r)}\in\mathbb{C}^{N_{R}\times 1}=[\mu_{l_{2},j,n}^{(r,1)},\ldots,\mu_{l_{2},j,n}^{(r,N_{R})}]^{T}, \nonumber\\
&\mu_{l_{2},j,n}^{(r,\bar{t})}=\sqrt{N_{R}N_{U}}e^{-\bar{j}2\pi\frac{j}{N_{S}T_{S}}\tau_{l_{2},n}}a_{R,n}^{*(\bar{t})}(\theta_{R,l_{2},j})a_{U,n}^{(r)}(\theta_{U,l_{2},j}),
 \mathbf{h}[t]\in\mathbb{C}^{N(L_{2}+1)\times 1}=[h_{1,0}[t],...,h_{N,L_{2}}[t]]^{T}.\label{formu18}
 \end{align}
Similarly, when channel matrix $\mathbf{H}_{n}[t,j]$ is known, the received signal is expressed as
\begin{align}
\mathbf{y}[t]=\mathbf{\Lambda}(\boldsymbol{\kappa}_{1},\boldsymbol{\theta},\mathbf{w},\mathbf{h}[t])\mathbf{g}[t]+\mathbf{n}[t],\label{formu19}
\end{align}
where $\tilde{t}\in\{1,\ldots,N_{T}\}$, $\boldsymbol{\kappa}_{1}\in\mathbb{R}^{(2N(L_{1}+1)+3)\times 1}=[\mathbf{x}^{T},\varphi,\mathbf{r}^{T}]$
\begin{align}
&\mathbf{\Lambda}(\boldsymbol{\kappa}_{1},\boldsymbol{\theta},\mathbf{w},\mathbf{h}[t])\in\mathbb{C}^{N_{U}N_{S}NM\times N(L_{1}+1)}=\left[\begin{matrix}
\boldsymbol{\Lambda}_{1}&\ldots&\mathbf{0}\\
\vdots&\ddots&\vdots\\
\mathbf{0}&\ldots&\boldsymbol{\Lambda}_{N}
  \end{matrix}
  \right],\nonumber\\
&\boldsymbol{\Lambda}_{n}\in\mathbb{C}^{N_{U}N_{S}M\times(L_{1}+1)}=[\bar{\boldsymbol{\gamma}}_{1}^{(r)}[1,1],\ldots,\bar{\boldsymbol{\gamma}}_{N}^{(r)}[N_{S},M]]^{T},\bar{\boldsymbol{\gamma}}_{n}^{(r)}[j,m]=[\bar{\gamma}_{0,n}^{(r)}[j,m],\ldots,\bar{\gamma}_{L_{1}+1,n}^{(r)}[j,m]]^{T},\nonumber\\
&\bar{\gamma}_{l_{1},n}^{(r)}[j,m]=\mathbf{w}_{n}^{H}[j,m]\bar{\boldsymbol{\mu}}_{l_{1},j,n}^{(r)},\bar{\boldsymbol{\mu}}_{l_{1},j,n}^{(r)}\in\mathbb{C}^{N_{T}\times 1}=[\bar{\mu}_{l_{1},j,n}^{(r,\tilde{t})},\ldots,\bar{\mu}_{l_{1},j,n}^{(r,\tilde{t})}]^{T},\nonumber\\
&\bar{\mu}_{l_{1},j,n}^{(r,\tilde{t})}=\sqrt{N_{R}N_{T}}e^{-\bar{j}2\pi\frac{j}{N_{S}T_{S}}\tau_{l_{1},n}}a_{T,n}^{*(\tilde{t})}(\theta_{T,l_{1},j})c^{r}_{n,j,l_{1}}[t],
\mathbf{B}_{n}[t,j]\in\mathbb{C}^{N_{U}\times N_{R}}=\mathbf{H}_{n}[t,j]\boldsymbol{\Theta},\nonumber\\
&\mathbf{c}_{n,j,l_{1}}[t]\in\mathbb{C}^{N_{U}\times 1}=\mathbf{B}_{n}[t,j]\mathbf{a}_{R,j}(\theta_{R,l_{1},n}),\mathbf{g}[t]\in\mathbb{C}^{N(L_{1}+1)\times 1}=[g_{1,0}[t],\ldots,g_{N,L_{1}}[t]]^{T},\label{formu23}
\end{align}
where $a_{U,n}^{(r)}(\theta_{U,l_{1},j})$ is the $r$-th element of $\mathbf{a}_{U,n}(\theta_{U,l_{1},j})$. $a_{R,n}^{(\bar{t})}(\theta_{R,l_{1},j})$ and $a_{\bar{R},n}^{(\bar{t})}(\theta_{\bar{R},l_{2},j})$ denote the $\bar{t}$-th element of $\mathbf{a}_{R,n}(\theta_{R,l_{1},j})$ and $\mathbf{a}_{\bar{R},n}(\theta_{\bar{R},l_{2},j})$.The above operation process is written in \textbf{Appendix~A}.
Note that the {location} parameters $\boldsymbol{\kappa}$ and beamforming vector are coupled. Therefore, the beamforming design and localization problem is non-convex mentioned above. To solve the non-convex problem, we propose a beamforming design scheme and explain how the algorithm deals with this problem. 
Next, we will elaborate on the principle and details of the proposed beamforming algorithm.

\section{Joint Localization and Beamforming Optimization Algorithm}\label{III}
In this paper, we need to estimate the transmit beamforming vector $\mathbf{w}$, the passive beamforming matrix of RIS $\mathbf{\Theta}$, and the {location} parameters of BS to RIS and RIS to UE $\boldsymbol{\kappa}$.
Based on the expression in (\ref{formu14}) and (\ref{formu19}), we can find that these parameters are coupled in expressions (\ref{formu14}) and (\ref{formu19}). According to \cite{6146494,6125211}, we propose the joint localization and beamforming optimization (JLBO) based on BCD algorithm to solve this non-convex problem. The details of the proposed JLBO algorithm based on BCD are given in \textbf{Algorithm~1}. 
\begin{algorithm}%
\caption{JLBO Algorithm Based on BCD} \label{algo3-1}%算法的名字
\hspace*{0.02in}{\bf Initialize:}
$(\mathbf{w}_{n}[j,m])^{(0)}$, $(\mathbf{g}[t])^{(0)}$,$(\mathbf{h}[t])^{(0)}$, $(\boldsymbol{\Theta})^{(0)}$ and $(\boldsymbol{\kappa})^{(0)}$,$i=0$.\\
\hspace*{0.02in}{\bf Repeat:}~$i=i+1$.\\
Given $(\mathbf{g}[t])^{(i-1)}$,$(\mathbf{w}_{n}[j,m])^{(i-1)}$, $(\boldsymbol{\Theta})^{(i-1)}$ and $(\boldsymbol{\kappa})^{(i-1)}$, {estimate} $(\mathbf{h}[t])^{(i)}$.\\
Given $(\mathbf{h}[t])^{(i)}$,$(\mathbf{w}_{n}[j,m])^{(i-1)}$, $(\boldsymbol{\Theta})^{(i-1)}$ and $(\boldsymbol{\kappa})^{(i-1)}$, {estimate} $(\mathbf{g}[t])^{(i)}$.\\ 
Given $(\mathbf{h}[t])^{(i)}$, $(\mathbf{g}[t])^{(i)}$ and $\boldsymbol{\kappa}^{(i-1)}$, {design} $(\mathbf{w}_{n}[j,m])^{(i)}$ and $(\boldsymbol{\Theta})^{(i)}$.\\ 
Given $(\mathbf{h}[t])^{(i)}$, $(\mathbf{g}[t])^{(i)}$, $(\mathbf{w}_{n}[j,m])^{(i)}$ and $(\boldsymbol{\Theta})^{(i)}$, {estimate} $\boldsymbol{\kappa}^{(i)}$.\\ 
\hspace*{0.02in}{\bf Until:}~Algorithm Convergence.\\
\hspace*{0.02in}{\bf Output:}
$\hat{\boldsymbol{\kappa}}$, $\hat{\mathbf{w}}_{n}[j,m]$, $\hat{\mathbf{h}}[t]$,$\hat{\mathbf{g}}[t]$ $\hat{\mathbf{\Theta}}$.\\
\end{algorithm}
Specifically, the JLBO algorithm based on BCD includes four parts, in the $3$-th step, we fix these parameters $\mathbf{g}[t]$, $\mathbf{w}$ and $\boldsymbol{\kappa}$, and the convex optimization algorithm based LS is used to estimate $\mathbf{h}[t]$. Similarly, in the $4$-th step, we fix variables $\mathbf{h}[t]$, $\mathbf{w}$, $\boldsymbol{\kappa}$ and use LS to estimate channel gain $\mathbf{g}[t]$. In the $5$-th step, transmit beamforming $\mathbf{w}_{n}[j,m]$ and passive beamforming of RIS $\boldsymbol{\Theta}$ is optimized based on CRLB. In the $6$-th, we use successive convex approximation (SCA) algorithm to estimate the
{location} parameters $\boldsymbol{\kappa}$ given other parameters. In the following, we will describe in detail the parameter estimation process in steps 3-6 in the \textbf{Algorithm~1}. 

\subsection{Channel gain estimation}
According to the receive signal model in (\ref{formu14}), channel gain $\mathbf{h}[t]$ estimation problem can be rewritten as
\begin{align}
\arg\min_{\mathbf{h}[t]}\|\mathbf{y}[t]-\boldsymbol{\Gamma}(\boldsymbol{\kappa}_{2},\boldsymbol{\theta},\mathbf{w},\mathbf{g}[t])\mathbf{h}[t]\|^{2},\label{formu24}
\end{align}
Since problem (\ref{formu24}) is convex with respect to (w.r.t.) the channel gain $\mathbf h[t]$, the optimal solution $\hat{\mathbf{h}}[t]$ can be estimated based on LS algorithm and  $\hat{\mathbf{h}}[t]$ is expressed as
\begin{align}
\hat{\mathbf{h}}[t]=(\boldsymbol{\Gamma}(\boldsymbol{\kappa}_{2},\boldsymbol{\theta},\mathbf{w},\mathbf{g}[t]))^{\dagger}\mathbf{y}[t].\label{formu25}
\end{align}
After $\mathbf{h}[t]$ is obtained, the received signal can be updated as $\hat{\mathbf{y}}[t]$. Similarly, the estimation problem of $\mathbf{g}[t]$ is given as
\begin{align}
\arg\min_{\mathbf{g}[t]}\|\hat{\mathbf{y}}[t]-\boldsymbol{\Lambda}(\boldsymbol{\kappa}_{1},\boldsymbol{\theta},\mathbf{w},{\color{red}\hat{\mathbf{h}}[t]})\mathbf{g}[t]\|^{2},\label{formu26}
\end{align}
whose optimal solution is expressed as
\begin{align}
\hat{\mathbf{g}}[t]=(\boldsymbol{\Lambda}(\boldsymbol{\kappa}_{1},\boldsymbol{\theta},\mathbf{w},\hat{\mathbf{h}}[t]))^{\dagger}\hat{\mathbf{y}}[t].\label{formu27}
\end{align}
Therefore, $(\mathbf{h}[t])^{(i)}=\hat{h}[t]$ and $(\mathbf{g}[t])^{(i)}=\hat{\mathbf{g}}[t]$ in $3$-th and $4$-th step in \textbf{Algorithm~1}.
\subsection{Transmit and RIS passive beamforming optimization}
The beamforming optimization problem can be rewritten as
\begin{align}
\arg\min_{\boldsymbol{\theta},\mathbf{w}}\|\mathbf{y}[t]-\boldsymbol{\Gamma}(\boldsymbol{\kappa}_{2},\boldsymbol{\theta},\mathbf{w},\mathbf{g}[t])\mathbf{h}[t]\|^{2},\label{formu34}\\
\arg\min_{\boldsymbol{\theta},\mathbf{w}}\|\mathbf{y}[t]-\boldsymbol{\Lambda}(\boldsymbol{\kappa}_{1},\boldsymbol{\theta},\mathbf{w},\mathbf{h}[t])\mathbf{g}[t]\|^{2}.\label{formu35}
\end{align}
We note that the MSE depends on beamforming vector $\mathbf{w}$ and passive beamforming vector $\boldsymbol{\theta}$. 
However, because these variables are coupled and complex, it is difficult to get a closed expression of the MSE. To solve this problem, we use {\color{red}CRLB} to deal with this problem instead of MSE for the objective function and optimize the transmit beamforming vector $\mathbf{w}$. 
According to \cite{6146494,6799306}, this method is effective because the decline of CRLB also means the decline of MSE. For the CRLB, we have $\textbf{Theorem~1}$.

\begin{theorem}\label{The31}
The CRLB of MSE is about $\mathbf{w}$ and $\boldsymbol{\theta}$, and the CRLB of (\ref{formu34}) and (\ref{formu35}) are denoted as,
\begin{align}
\mathrm{MSE}(\mathbf{w},\boldsymbol{\theta},\mathbf{g}[t])\geq \underbrace{\mathrm{trace}(\boldsymbol{\mathcal{J}}(\boldsymbol{\kappa}_{2},\mathbf{h}[t];\mathbf{w},\boldsymbol{\theta},\mathbf{g}[t])^{-1})}_{\mathrm{CRLB}_{1}(\mathbf{w},\boldsymbol{\theta},\mathbf{g}[t])},\nonumber\\
\mathrm{MSE}(\mathbf{w},\boldsymbol{\theta},\mathbf{h}[t])\geq\underbrace{\mathrm{trace}(\boldsymbol{\mathcal{I}}(\boldsymbol{\kappa}_{1},\mathbf{g}[t];\mathbf{w},\boldsymbol{\theta},\mathbf{h}[t])^{-1})}_{\mathrm{CRLB}_{2}(\mathbf{w},\boldsymbol{\theta},\mathbf{h}[t])}.\label{formu37}
\end{align}
The proof is given in \textbf{Appendix~B}.
\end{theorem}
According to $\textbf{Theorem~1}$, MSE depends on $\mathbf{w}$, $\boldsymbol{\theta}$,  $\mathbf{g}[t]$ and $\mathbf{h}[t]$, since $\mathbf{g}[t]$ and $\mathbf{h}[t]$ can be obtained based on (\ref{formu25}) and (\ref{formu27}) The CRLB of MSE is rewritten as
\begin{align}
\mathrm{CRLB}(\mathbf{w},\boldsymbol{\theta})=\mathrm{CRLB}_{1}(\mathbf{w},\boldsymbol{\theta},\hat{\mathbf{g}}[t])+\mathrm{CRLB}_{2}(\mathbf{w},\boldsymbol{\theta},\hat{\mathbf{h}}[t]).\label{formu38}
\end{align}
Therefore, we propose to design the beamforming vector $\mathbf{w}$ and RIS reflecting vector $\boldsymbol{\theta}$ as the following minimization problem
\begin{subequations}
\begin{align}
\min_{\mathbf{w},\boldsymbol{\theta}}&~\mathrm{CRLB}(\mathbf{w},\boldsymbol{\theta})\label{formu39}\\
\mbox{s.t.}~
&\|\mathbf{w}_{n}[j,m]\|^{2}\leq 1, \forall n,m,j,&\label{formu40}\\
&|\boldsymbol{\theta}_{i}|=1, 1\leq i\leq N_{r},&\label{formu41}
\end{align}\label{formu42}%
\end{subequations}
where (\ref{formu40}) and (\ref{formu41}) are the transmit power constraint and the RIS phase shifts constraint, respectively. 
In this part, we adopt the alternating optimization strategy. First, we fix the passive beamforming matrix $\boldsymbol{\theta}$, using SCA algorithm to optimize the transmit beamforming vector $\mathbf{w}$. After $\mathbf{w}$ is given, we continue to use SCA algorithm to obtain $\boldsymbol{\theta}$. Thus, problem (\ref{formu42}) can be divided into two problems and the transmit beamforming optimization problem is expressed as
\begin{subequations}
\begin{align}
\min_{\mathbf{w}}&~\mathrm{CRLB}(\mathbf{w},\boldsymbol{\theta})\label{formu43}\\
\mbox{s.t.}~
&\|\mathbf{w}_{n}[j,m]\|^{2}\leq 1, \forall n,m,j.&\label{formu44}
\end{align}\label{formu45}%   
\end{subequations}
To address the non-convex problem in (\ref{formu43}), a non-convex gradient descent optimization algorithm based on SCA is proposed in this section. The surrogate function of (\ref{formu43}) at $i+1$-th iteration is denoted as $\bar{f}_{S}(\mathbf{w};\mathbf{w}^{(i)})$ as follows
\begin{align}
&\bar{f}_{S}(\mathbf{w};\mathbf{w}^{(i)})=C(\boldsymbol{\kappa};\mathbf{w}^{(i)})-\nonumber\\
&\sum\limits_{j,n,m}\mathbf{w}_{n}^{T}[j,m]\mathbf{G}_{n}^{T}[t,j]\boldsymbol{\Theta}^{T}\left(\sum_{l_{2},r}\frac{\sigma_{l_{2}}^{2}\mathbf{U}_{l_{2},j,n}^{(r)H}(\boldsymbol{\kappa}_{2})\mathbf{B}^{H}(\mathbf{w}^{(i)})\mathbf{B}(\mathbf{w}^{(i)})\mathbf{U}_{l_{2},j,n}^{(r)}(\boldsymbol{\kappa}_{2})}{\sigma^{2}}\right)\boldsymbol{\Theta}^{*}\mathbf{G}_{n}^{*}[t,j]\mathbf{w}_{n}^{*}[j,m]-\nonumber\\
&\sum\limits_{j,n,m}\mathbf{w}_{n}^{T}[j,m]\mathbf{G}_{n}^{T}[t,j]\boldsymbol{\Theta}^{T}\left(\sum_{l_{2},r}\frac{\mathbf{V}_{j,n}^{(r)}(\boldsymbol{\kappa}_{2})\mathbf{E}^{H}(\mathbf{w}^{(i)})\mathbf{E}(\mathbf{w}^{(i)})\mathbf{V}_{j,n}^{(r)H}(\boldsymbol{\kappa}_{2})}{\sigma^{2}}\right)\boldsymbol{\Theta}^{*}\mathbf{G}_{n}^{*}[t,j]\mathbf{w}_{n}^{*}[j,m]-\nonumber\\
&\sum\limits_{j,n,m}\mathbf{w}_{n}^{T}[j,m]\left(\sum_{l_{1},r}\frac{\sigma_{l_{1}}^{2}\bar{\mathbf{U}}_{l_{1},j,n}^{(r)H}(\boldsymbol{\kappa}_{1})\bar{\mathbf{B}}^{H}(\mathbf{w}^{(i)})\bar{\mathbf{B}}(\mathbf{w}^{(i)})\bar{\mathbf{U}}_{l_{1},j,n}^{(r)}(\boldsymbol{\kappa}_{1})}{\sigma^{2}}\right)\mathbf{w}_{n}^{*}[j,m]-\nonumber\\
&\sum\limits_{j,n,m}\mathbf{w}_{n}^{T}[j,m]\left(\sum_{l_{1},r}\frac{\bar{\mathbf{V}}_{j,n}^{(r)H}(\boldsymbol{\kappa}_{1})\bar{\mathbf{E}}^{H}(\mathbf{w}^{(i)})\bar{\mathbf{E}}(\mathbf{w}^{(i)})\bar{\mathbf{V}}_{j,n}^{(r)}(\boldsymbol{\kappa}_{1})}{\sigma^{2}}\right)\mathbf{w}_{n}^{*}[j,m],\label{formu46}
\end{align}
where
\begin{align}
&\mathbf{U}_{l_{2},j,n}^{(r)}(\boldsymbol{\kappa}_{2})\in\mathbb{C}^{(2N(L_{2}+1)+3)\times N_{R}}=\left[\begin{matrix}
\mathbf{u}_{l_{2},j,n}^{(r,1)},\ldots,
\mathbf{u}_{l_{2},j,n}^{(r,N_{R})}
  \end{matrix}
  \right]^{T},\nonumber\\
&\mathbf{V}_{j,n}^{(r)}(\boldsymbol{\kappa}_{2})\in\mathbb{C}^{N_{R}\times (L_{1}+1)}=\left[\begin{matrix}
\sum_{\tilde{t}=1}^{N_{T}}\mathbf{\zeta}_{0,j,n}^{(r,\tilde{t})},\ldots,
\sum_{\tilde{t}=1}^{N_{T}}\mathbf{\zeta}_{L_{2},j,n}^{(\tilde{t})}
  \end{matrix}
  \right]^{T},\boldsymbol{\zeta}_{l_{2},j,n}^{r,\tilde{t},\bar{t}}\in\mathbb{C}^{N_{R}\times 1}=\mathbf{G}_{n}[t,j]^{H}\boldsymbol{\Theta}^{H}\boldsymbol{\mu}_{l_{2},j,n}^{(r)},\nonumber\\
&\mathbf{u}_{l_{2},j,n}^{(r,\bar{t})}\in\mathbb{C}^{(2N(L_{2}+1)+3)\times 1}=\left[\begin{matrix}
\sum_{\tilde{t}=1}^{N_{T}}\zeta_{l_{2},j,n}^{r,\tilde{t}}\mathbf{p}_{l_{2},j,n}^{r,\bar{t}},\sum_{\tilde{t}=1}^{N_{T}}\zeta_{l_{2},j,n}^{r,\tilde{t}}\mathbf{o}_{l_{2},j,n}^{r,\bar{t}},\sum_{\tilde{t}=1}^{N_{T}}\zeta_{l_{2},j,n}^{r,\tilde{t}\mathbf{q}_{l_{2},j,n}^{r,\bar{t}}}
\end{matrix}
\right]^{T},\nonumber\\
&\mathbf{B}(\mathbf{w}^{(i)})=(\sum_{l_{2},j,r,n,m}\eta_{l}^{2}\mathbf{U}_{l,j,n}^{(r)}(\boldsymbol{\kappa}_{2})\boldsymbol{\Theta}^{*}\mathbf{G}_{n}^{*}[t,j]\mathbf{w}^{*(i)}_{n}[j,m]\mathbf{w}^{T(i)}_{n}[j,m]\mathbf{G}_{n}^{T}[t,j]\boldsymbol{\Theta}^{T}\mathbf{U}_{l_{2},j,n}^{(r)H}(\boldsymbol{\kappa}_{2}))^{-1},\nonumber\\
&\mathbf{E}(\mathbf{w}^{(i)})=(\sum_{r,j,m}\mathbf{V}_{j,n}^{(r)}(\boldsymbol{\kappa}_{2})\boldsymbol{\Theta}^{*}\mathbf{G}_{n}^{*}[t,j]\mathbf{w}^{*(i)}_{n}[j,m]\mathbf{w}^{T(i)}_{n}[j,m]\mathbf{G}_{n}^{T}[t,j]\boldsymbol{\Theta}^{T}\mathbf{V}_{j,n}^{(r)H}(\boldsymbol{\kappa}_{2}))^{-1},\nonumber\\
&\bar{\mathbf{U}}_{l_{1},j,n}^{(r,t)}\in\mathbb{C}^{(2N(L_{1}+1)+3)\times N_{T}}=\left[\begin{matrix}
\bar{\mathbf{u}}_{l_{1},j,n}^{(r,1)},
\ldots,
\bar{\mathbf{u}}_{l_{1},j,n}^{(r,N_{T})}
  \end{matrix}
  \right]^{T},\bar{\mathbf{V}}_{j,n}^{(r)}\in\mathbb{C}^{(2N(L_{1}+1)+3)\times N_{T}}=\left[\begin{matrix}
\bar{\boldsymbol{\mu}}_{0,j,n}^{(r)},
\ldots,
\bar{\boldsymbol{\mu}}_{L_{1},j,n}^{(r)}
  \end{matrix}
  \right]^{T},\nonumber\\
&\bar{\mathbf{u}}_{l_{1},j,n}^{(r,\tilde{t})}\in\mathbb{C}^{(2N(L_{1}+1)+3)\times 1}=\left[\begin{matrix}
\bar{\mu}_{l_{1},j,n}^{r,\tilde{t}}\bar{\mathbf{p}}_{l_{1},j,n}^{r,\tilde{t}},\bar{\mu}_{l_{1},j,n}^{r,\tilde{t}}\mathbf{o}_{l_{2},j,n}^{r,\tilde{t}},\bar{\mu}_{l_{1},j,n}^{r,\tilde{t}}\mathbf{q}_{l_{2},j,n}^{r,\tilde{t}}
\end{matrix}
\right]^{T},\nonumber\\
&\bar{\mathbf{B}}(\mathbf{w}^{(i)})=(\sum_{l_{1},j,r,n,m}\eta_{l}^{2}\bar{\mathbf{U}}_{l_{1},j,n}^{(r)}(\boldsymbol{\kappa}_{1})\mathbf{w}^{*(i)}_{n}[j,m]\mathbf{w}^{T(i)}_{n}[j,m]\bar{\mathbf{U}}_{l_{1},j,n}^{(r)H}(\boldsymbol{\kappa}_{1}))^{-1},\nonumber\\
&\bar{\mathbf{E}}(\mathbf{w}^{(i)})=(\sum_{r,j,m}\bar{\mathbf{V}}_{j,n}^{(r)}(\boldsymbol{\kappa}_{1})\mathbf{w}^{*(i)}_{n}[j,m]\mathbf{w}^{T(i)}_{n}[j,m]\bar{\mathbf{V}}_{j,n}^{(r)H}(\boldsymbol{\kappa}_{1}))^{-1}.\label{formu52}
\end{align}
These variables of (\ref{formu52}) is given in \textbf{Appendix~B}.
The surrogate function $\bar{f}_{S}(\mathbf{w};\mathbf{w}^{(i)})$ in (\ref{formu46}) is based on the first-order Taylor expansion of $\bar{f}_{S}(\mathbf{w};\mathbf{w}^{(i)})$. Note that we have used SCA method to obtain the surrogate function, but the problem in (\ref{formu46}) is still a non-convex problem. The surrogate function is a better approximation of $\mathrm{CRLB}(\mathbf{w},\boldsymbol{\theta})$ with faster convergence. Moreover, compared with the original objective function $\mathrm{CRLB}(\mathbf{w},\boldsymbol{\theta})$, a closed-form gradient vector can be obtained by the surrogate function. Thus, using the surrogate function will lead to a lower computational cost.
According to the expression in (\ref{formu46}), after transmit beamforming vector $\mathbf{w}^{(i)}$ at $i$-th iteration is given, the update problem of $\mathbf{w}^{(i)}$ is rewritten as
\begin{subequations}
\begin{align}
\mathbf{w}^{(i+1)}=\arg\min_{\mathbf{w}}&~\bar{f}_{S}(\mathbf{w};\mathbf{w}^{(i)})\label{formu53}\\
\mbox{s.t.}~
&\|\mathbf{w}_{n}[j,m]\|^{2}\leq 1.&\label{formu54}
\end{align}\label{formu55}%
\end{subequations}
Since the problem in (\ref{formu55}) is still non-convex, we can use the gradient descent search algorithm to obtain the local optimal solution of the non-convex problem. Based on Rayleigh quotient maximization in\cite{zhang2013optimizing}, problem (\ref{formu55}) can be rewritten as
\begin{align}
\mathbf{w}^{(i+1)}=\arg\min_{\mathbf{w}}&~\frac{\bar{f}_{S}(\mathbf{w};\mathbf{w}^{(i)})}{\|\mathbf{w}^{(i)}\|^{2}}.\label{formu57}
\end{align}
Therefore, to maximize the objective cost function, $\mathbf{w}^{(i+1)}$ is obtained based on the principal eigenvector of $\mathbf{B}(\mathbf{w}^{(i)})$. Once $\mathbf{d}^{(i+1)}$ in (\ref{formu55}) is determined, according to the Armijo rule \cite{bertsekas1997nonlinear,8752072}, the transmit beamforming vector is expressed as
\begin{align}
\mathbf{w}_{n}^{(i)}[j,m]=\mathbf{w}_{n}^{(i)}[j,m]+\bar{\lambda}^{(i)}\mathbf{d}_{n}^{(i+1)}[j,m], \label{formu58}   
\end{align}
{\color{red}where the update direction vector $\mathbf{d}_{n}^{(i+1)}[j,m]$ is expressed as}
\begin{align}
\mathbf{d}_{n}^{(i+1)}[j,m]=\mathbf{w}_{n}^{(i+1)}[j,m]-\mathbf{w}_{n}^{(i)}[j,m].\label{formu56}
\end{align}
$\bar{\lambda}^{(i)}>0$ is the step length determined by the Armijo rule in \cite{bertsekas1997nonlinear,8752072} in (\ref{formu58}) shown
\begin{align}
&\bar{f}_{S}(\mathbf{w}_{n}^{(i)}[j,m]+\bar{\lambda}^{(i)}\mathbf{d}_{n}^{(i+1)}[j,m])\leq \bar{f}_{S}(\mathbf{w}_{n}^{(i)}[j,m])+\nonumber\\
&a\bar{\lambda}\mathcal{R}\{\nabla_{\mathbf{w}_{n}^{(i)}[j,m]}\bar{f}_{S}(\mathbf{w}_{n}^{(i)}[j,m])\mathbf{d}_{n}^{(i+1)}[j,m]\},\label{formu59}
\end{align}
where $\nabla_{\mathbf{w}_{n}^{(i)}[j,m]}\bar{f}_{S}(\mathbf{w}_{n}^{(i)}[j,m])$ is expressed as
\begin{align}
\nabla_{\mathbf{w}_{n}^{(i)}[j,m]}\bar{f}_{S}(\mathbf{w}_{n}^{(i)}[j,m])=-2\frac{(\|\mathbf{w}_{n}^{(i)}[j,m]\|^{2}\mathbf{I}_{N_{T}}-\mathbf{w}_{n}^{*(i)}[j,m]\mathbf{w}_{n}^{T(i)}[j,m])}{\|\mathbf{w}_{n}^{(i)}[j,m]\|^{4}}\mathbf{B}(\mathbf{w}^{(i)})\mathbf{w}_{n}^{*(i)}[j,m].\label{formu60}
\end{align}
Specifically, starting with a certain step size $\bar{\lambda}>0$, the Armijo rule repeatedly decrease $\bar{\lambda}$ as $\bar{\lambda}=\bar{l}\ \bar{\lambda}$ for some $\bar{l}\in (0,1)$ until the condition in (\ref{formu58}) is satisfied. Similarly, the RIS reflection vector $\boldsymbol{\theta}$ optimization problem can be written as
\begin{subequations}
\begin{align}
\min_{\boldsymbol{\theta}}&~\mathrm{CRLB}(\mathbf{w},\boldsymbol{\theta})\label{formu61}\\
\mbox{s.t.}~
&|\boldsymbol{\theta}_{i}|=1, 1\leq i\leq N_{R}.&\label{formu62}
\end{align}\label{formu63}%
\end{subequations}
Due to the problem in (\ref{formu63}) is non-convex, we use the penalty factor method to substitute the constraint condition in (\ref{formu62}) into objective function. Thus, (\ref{formu63}) is rewritten as
\begin{align}
\min_{\boldsymbol{\theta}}&~\mathrm{CRLB}(\mathbf{w},\boldsymbol{\theta})+\eta\sum_{i=1}^{N_{R}}(|\boldsymbol{\theta}_{i}|-1)^{2},\label{formu64}
\end{align}
where $\eta>0$ is enough large. 
For the objective function in (\ref{formu64}), {\color{red}its surrogate} function at $(i+1)$-th iteration is denoted as $\tilde{f}_{S}(\boldsymbol{\theta};\boldsymbol{\theta}^{(i)})$, which is written as
\begin{align}
&\bar{f}_{S}(\boldsymbol{\theta};\boldsymbol{\theta}^{(i)})=C(\boldsymbol{\kappa};\boldsymbol{\theta}^{(i)})+\eta\sum_{i=1}^{N_{R}}\boldsymbol{\theta}_{i}^{(i)}(\boldsymbol{\theta}_{i}-\boldsymbol{\theta}_{i}^{(i)})-\nonumber\\
&\sum_{j,n,m}\boldsymbol{\theta}^{T}\left(\sum_{l_{2},r}\frac{\sigma_{l_{2}}^{2}\check{\mathbf{U}}_{l_{2},j,n}^{(r)H}(\boldsymbol{\kappa}_{2})\check{\mathbf{B}}^{H}(\boldsymbol{\theta})\check{\mathbf{B}}(\boldsymbol{\theta})\check{\mathbf{U}}_{l_{2},j,n}^{(r)}(\boldsymbol{\kappa})(\boldsymbol{\kappa}_{2})}{\sigma^{2}}\right)\boldsymbol{\theta}^{*}-\nonumber\\
&\sum_{j,n,m}\boldsymbol{\theta}^{T}\left(\sum_{l_{2},r}\frac{\sigma_{l_{2}}^{2}\check{\mathbf{V}}_{j,n}^{(r)}(\boldsymbol{\kappa}_{2})\check{\mathbf{E}}^{H}(\boldsymbol{\theta})\check{\mathbf{E}}(\boldsymbol{\theta})\check{\mathbf{V}}_{j,n}^{(r)H}(\boldsymbol{\kappa}_{2})}{\sigma^{2}}\right)\boldsymbol{\theta}^{*}-\nonumber\\
&\sum\limits_{j,n,m}\boldsymbol{\theta}^{T}\left(\sum_{l_{1},r}\frac{\sigma_{l_{1}}^{2}\breve{\mathbf{U}}_{l_{1},j,n}^{(r)H}(\boldsymbol{\kappa}_{1})\breve{\mathbf{B}}^{H}(\boldsymbol{\theta}^{(i)})\breve{\mathbf{B}}(\boldsymbol{\theta}^{(i)})\breve{\mathbf{U}}_{l_{1},j,n}^{(r)}(\boldsymbol{\kappa}_{1})}{\sigma^{2}}\right)\boldsymbol{\theta}^{*}-\nonumber\\
&\sum\limits_{j,n,m}\boldsymbol{\theta}^{T}\left(\sum_{l_{1},r}\frac{\breve{\mathbf{V}}_{j,n}^{(r)}(\boldsymbol{\kappa}_{1})\breve{\mathbf{E}}^{H}(\boldsymbol{\theta}^{(i)})\breve{\mathbf{E}}(\boldsymbol{\theta}^{(i)})\breve{\mathbf{V}}_{j,n}^{(r)H}(\boldsymbol{\kappa}_{1})}{\sigma^{2}}\right)\boldsymbol{\theta}^{*},\label{formu65}
\end{align}
where 
\begin{align}
&\check{\mathbf{U}}_{l_{2},j,n}^{(r)}\in\mathbb{C}^{(2N(L_{2}+1)+3)\times N_{R}}=\left[\begin{matrix}
\check{\mathbf{u}}_{l_{2},j,n}^{(r,1)},\ldots,
\check{\mathbf{u}}_{l_{2},j,n}^{(r,N_{R})}
  \end{matrix}
  \right]^{T},\check{\mathbf{V}}_{j,n}^{(r)}\in\mathbb{C}^{N_{R}\times (L_{1}+1)}=\left[\begin{matrix}
\check{\mathbf{\zeta}}_{0,j,n}^{(r,\tilde{t})},\ldots,
\check{\mathbf{\zeta}}_{L_{2},j,n}^{(\tilde{t})}
  \end{matrix}
  \right]^{T},\nonumber\\
&\check{\mathbf{u}}_{l_{2},j,n}^{(r,\bar{t})}\in\mathbb{C}^{(2N(L_{2}+1)+3)\times 1}=\left[\begin{matrix}
\check{\zeta}_{l_{2},j,n}^{(r,\bar{t})}\mathbf{p}_{l_{2},j,n}^{(r,\bar{t})},\check{\zeta}_{l_{2},j,n}^{(r,\bar{t})}\mathbf{o}_{l_{2},j,n}^{(r,\bar{t})},\check{\eta}_{l_{2},j,n}^{(r,\bar{t})}\mathbf{q}_{l_{2},j,n}^{r,\bar{t}}
\end{matrix}
\right]^{T},\nonumber\\
&\check{\boldsymbol{\zeta}}_{l_{2},j,n}^{(r,\bar{t})}\in\mathbb{C}^{1\times N_{R}}=\mathbf{w}_{n}[j,m]^{H}\mathbf{G}_{n}[t,j]^{H}\mathrm{diag}(\boldsymbol{\mu}_{l_{2},j,n}^{(r)})\nonumber\\
&\mathbf{B}(\boldsymbol{\theta}^{(i)})=(\sum_{l_{2},j,r,n,m}\sigma_{l_{2}}^{2}\check{\mathbf{U}}_{l_{2},j,n}^{(r)}(\boldsymbol{\kappa}_{2})\mathrm{diag}(\mathbf{G}_{n}^{*}[t,j]\mathbf{w}^{*(i)}_{n}[j,m])\boldsymbol{\theta}^{*}\boldsymbol{\theta}^{T}\mathrm{diag}(\mathbf{w}^{T(i)}_{n}[j,m]\mathbf{G}_{n}^{T}[t,j])\check{\mathbf{U}}_{l,j,n}^{(r)H}(\boldsymbol{\kappa}_{2}))^{-1},\nonumber\\
&\mathbf{E}(\boldsymbol{\theta}^{(i)})=(\sum_{r,j,m}\check{\mathbf{V}}_{j,n}^{(r)}(\boldsymbol{\kappa}_{2})\mathrm{diag}(\mathbf{G}_{n}^{*}[t,j]\mathbf{w}^{*(i)}_{n}[j,m])\boldsymbol{\theta}^{*}\boldsymbol{\theta}^{T}\mathrm{diag}(\mathbf{w}^{T(i)}_{n}[j,m]\mathbf{G}_{n}^{T}[t,j])\check{\mathbf{V}}_{j,n}^{(r)H}(\boldsymbol{\kappa}_{2}))^{-1},\nonumber\\
&\breve{\mathbf{U}}_{l_{1},j,n}^{(r)}\in\mathbb{C}^{(2N(L_{1}+1)+3)\times N_{R}}=\left[\begin{matrix}
\breve{\mathbf{u}}_{l_{1},j,n}^{(r,1)},\ldots,
\breve{\mathbf{u}}_{l_{1},j,n}^{(r,N_{R})}
  \end{matrix}
  \right]^{T},\breve{\mathbf{V}}_{j,n}^{(r)}\in\mathbb{C}^{N_{R}\times (L_{1}+1)}=\left[\begin{matrix}
\sum_{\tilde{t}=1}^{N_{T}}\breve{\mathbf{\zeta}}_{0,j,n}^{(r,\tilde{t})},\ldots,
\sum_{\tilde{t}=1}^{N_{T}}\breve{\mathbf{\zeta}}_{L_{2},j,n}^{(\tilde{t})}
  \end{matrix}
  \right]^{T},\nonumber\\
&\breve{\mathbf{u}}_{l_{1},j,n}^{(r,\bar{t})}\in\mathbb{C}^{(2N(L_{1}+1)+3)\times 1}=\left[\begin{matrix}
\sum_{\tilde{t}=1}^{N_{T}}\breve{\zeta}_{l_{1},j,n}^{r,\tilde{t}}\bar{\mathbf{p}}_{l_{1},j,n}^{r,\bar{t}},\sum_{\tilde{t}=1}^{N_{T}}\breve{\zeta}_{l_{1},j,n}^{r,\tilde{t}}\bar{o}_{l_{1},j,n}^{r,\bar{t}},\sum_{\tilde{t}=1}^{N_{T}}\breve{\zeta}_{l_{1},j,n}^{r,\tilde{t}\bar{\mathbf{q}}_{l_{1},j,n}^{r,\bar{t}}}
\end{matrix}
\right]^{T},\nonumber\\
&\breve{\boldsymbol{\zeta}}_{l_{1},j,n}^{r,\tilde{t},\bar{t}}\in\mathbb{C}^{N_{R}\times 1}=\mathbf{H}_{n}[t,j]\mathrm{diag}(\mathbf{G}_{n}[t,j]^{H}\bar{\boldsymbol{\mu}}_{l_{1},j,n}^{(r)})\mathbf{w}_{n}[t,j]\nonumber\\
&\breve{\mathbf{B}}(\boldsymbol{\theta}^{(i)})=(\sum_{l_{1},j,r,n,m}\sigma_{l_{1}}^{2}\breve{\mathbf{U}}_{l_{1},j,n}^{(r)}(\boldsymbol{\kappa}_{1})\mathrm{diag}(\mathbf{G}_{n}^{*}[t,j]\mathbf{w}^{*(i)}_{n}[j,m])\boldsymbol{\theta}^{(i)*}\boldsymbol{\theta}^{(i)T}\mathrm{diag}(\mathbf{w}^{T(i)}_{n}[j,m]\mathbf{G}_{n}^{T}[t,j])\mathbf{U}_{l_{1},j,n}^{(r)H}(\boldsymbol{\kappa}_{1}))^{-1},\nonumber\\
&\breve{\mathbf{E}}(\boldsymbol{\theta}^{(i)})=(\sum_{r,j,m}\breve{\mathbf{V}}_{j,n}^{(r)}(\boldsymbol{\kappa}_{1})\mathrm{diag}(\mathbf{G}_{n}^{*}[t,j]\mathbf{w}^{*(i)}_{n}[j,m])\boldsymbol{\theta}^{(i)*}\boldsymbol{\theta}^{(i)T}\mathrm{diag}(\mathbf{w}^{T(i)}_{n}[j,m]\mathbf{G}_{n}^{T}[t,j])\breve{\mathbf{V}}_{j,n}^{(r)H}(\boldsymbol{\kappa}_{1}))^{-1},
\label{formu68}
\end{align}
The optimization steps of $\boldsymbol{\theta}^{(i+1)}$ are the same as those of $\mathbf{w}^{(i+1)}$. In order to save space, we omit these same steps.

\subsection{{Location} Parameter Estimation}
After $\mathbf{h}[t]$, $\mathbf{g}[t]$, $\mathbf{w}$ and $\boldsymbol{\theta}$ are obtained, we update the received signal $\mathbf{y}[t]$. Hence, the {location} parameters estimation problem is written as
\begin{align}
\arg\min_{\boldsymbol{\kappa}_{2}}\|\mathbf{y}[t]-\boldsymbol{\Gamma}(\boldsymbol{\kappa}_{2},\boldsymbol{\theta},\mathbf{w},\mathbf{g}[t])\mathbf{h}[t]\|^{2},\arg\min_{\boldsymbol{\kappa}_{1}}\|\mathbf{y}[t]-\boldsymbol{\Lambda}(\boldsymbol{\kappa}_{1},\boldsymbol{\theta},\mathbf{w},\mathbf{h}[t])\mathbf{g}[t]\|^{2}.\label{formu28}
\end{align}
According to the expression in (\ref{formu28}), {location} parameters $\boldsymbol{\kappa}_{2}$ are coupled, thus, the {location} parameters estimation problem is non-convex problem. To find the solution of $\boldsymbol{\kappa}_{2}$, we use SCA method to approximate the left-hand cost function in (\ref{formu28}) and the convex approximation function is expressed as  
\begin{align}
f_{S}(\boldsymbol{\kappa}_{2};\boldsymbol{\kappa}_{2}^{(i)},\hat{\mathbf{g}}[t],\hat{\mathbf{h}}[t])=\|\mathbf{y}[t]-\boldsymbol{\Gamma}(\boldsymbol{\kappa}_{2}^{(i)},\boldsymbol{\theta},\mathbf{w},\mathbf{g}[t])\mathbf{h}[t]-\nonumber\\
\nabla_{\boldsymbol{\kappa}_{2}}^{H}(\boldsymbol{\Gamma}(\boldsymbol{\kappa}_{2}^{(i)},\boldsymbol{\theta},\mathbf{w},\mathbf{g}[t])\mathbf{h}[t])(\boldsymbol{\kappa}_{2}-\boldsymbol{\kappa}_{2}^{(i)})\|^{2},\label{formu29}
\end{align}
where $\nabla_{\boldsymbol{\kappa}_{2}}^{H}(\boldsymbol{\Gamma}(\boldsymbol{\kappa}_{2}^{(i)},\boldsymbol{\theta},\mathbf{w},\mathbf{g}[t])\in\mathbb{C}^{N_{U}N_{S}NM\times (2N(L_{2}+1)+3)}$ and $\nabla_{\boldsymbol{\kappa}_{2}}(\boldsymbol{\Gamma}(\boldsymbol{\kappa}_{2}^{(i)},\boldsymbol{\theta},\mathbf{w},\mathbf{g}[t])$ is expressed as
\begin{align}
\left[\begin{matrix}
\sqrt{N_{T}N_{U}}(\mathbf{1}_{M}\otimes(\mu_{0,1,1}^{(1,1)*}\mathbf{p}_{0,1,1}^{(1,1)})^{H})^{H}\mathrm{diag}((\mathbf{I}_{N_{T}}\otimes\mathrm{diag}[h_{0:L_{2},1,(i)}[t]])(\mathbf{w}_{n}[t,1,1]\otimes\mathbf{1}_{L_{2}+1}))^{T}\\
\vdots\\
\sqrt{N_{T}N_{U}}(\mathbf{1}_{M}\otimes(\mu_{L_{2},N_{S},N}^{(N_{U},N_{T})*}\mathbf{p}_{L_{2},N_{S},N}^{(N_{U},N_{T})})^{H})^{H}\mathrm{diag}((\mathbf{I}_{N_{T}}\otimes\mathrm{diag}[h_{0:L_{2},N,(i)}[t]])(\mathbf{w}_{n}[t,N_{S},M]\otimes\mathbf{1}_{L_{2}+1}))^{T}\\
\sqrt{N_{T}N_{U}}(\mathbf{1}_{M}\otimes(\mu_{0,1,1}^{(1,1)*}o_{0,1,1}^{(1,1)})^{H})^{H}\mathrm{diag}((\mathbf{I}_{N_{T}}\otimes\mathrm{diag}[h_{0:L_{2},1,(i)}[t]])(\mathbf{w}_{n}[t,1,1]\otimes\mathbf{1}_{L_{2}+1}))^{T}\\
\vdots\\
\sqrt{N_{T}N_{U}}(\mathbf{1}_{M}\otimes(\mu_{L_{2},N_{S},N}^{(N_{U},N_{T})*}o_{L_{2},N_{S},N}^{(N_{U},N_{T})})^{H})^{H}\mathrm{diag}((\mathbf{I}_{N_{T}}\otimes\mathrm{diag}[h_{0:L_{2},N,(i)}[t]])(\mathbf{w}_{n}[t,N_{S},M]\otimes\mathbf{1}_{L_{2}+1}))^{T}\\
\sqrt{N_{T}N_{U}}(\mathbf{1}_{M}\otimes(\mu_{0,1,1}^{(1,1)*}\mathbf{q}_{0,1,1:N}^{(1,1)})^{H})^{H}\mathrm{diag}((\mathbf{I}_{N_{T}}\otimes\mathrm{diag}[h_{0:L_{2},1,(i)}[t]])(\mathbf{w}_{n}[t,1,1]\otimes\mathbf{1}_{L_{2}+1}))^{T}\\
\vdots\\
\sqrt{N_{T}N_{U}}(\mathbf{1}_{M}\otimes(\mu_{L_{2},N_{S},N}^{(N_{U},N_{T})*}\mathbf{q}_{L_{2},N_{S},1:N}^{(N_{U},N_{T})})^{H})^{H}\mathrm{diag}((\mathbf{I}_{N_{T}}\otimes\mathrm{diag}[h_{0:L_{2},N,(i)}[t]])(\mathbf{w}_{n}[t,N_{S},M]\otimes\mathbf{1}_{L_{2}+1}))^{T}
  \end{matrix}
  \right].\label{formu30}
\end{align}
Then, we iteratively solve the convex subproblems to obtain {location} parameters
\begin{align}
\boldsymbol{\kappa}_{2}^{(i+1)}=\arg\min_{\boldsymbol{\kappa}_{2}}f_{S}(\boldsymbol{\kappa}_{2};\boldsymbol{\kappa}_{2}^{(i)},\mathbf{g}[t],\mathbf{h}[t]), \label{formu31} 
\end{align}
Note that $f_{S}(\boldsymbol{\kappa}_{2};\boldsymbol{\kappa}_{2}^{(i)},\mathbf{g}[t],\mathbf{h}[t])$ is strictly convex w.r.t. variable $\boldsymbol{\kappa}_{2}$, hence, we can give the optimal solution of $\boldsymbol{\kappa}_{2}$ and localization variable $\boldsymbol{\kappa}_{2}^{(i+1)}$ of the $i+1$-th iteration is given
\begin{align}
\boldsymbol{\kappa}_{2}^{(i+1)}=\boldsymbol{\kappa}_{2}^{(i)}+(\nabla_{\boldsymbol{\kappa}_{2}}^{H}(\boldsymbol{\Gamma}(\boldsymbol{\kappa}_{2}^{(i)},\boldsymbol{\theta},\mathbf{w},\mathbf{g}[t])\mathbf{h}[t]))^{\dagger}(\mathbf{y}[t]-\boldsymbol{\Gamma}(\boldsymbol{\kappa}_{2}^{(i)},\boldsymbol{\theta},\mathbf{w},\mathbf{g}[t])\mathbf{h}[t]).\label{formu32}
\end{align}
The second term on the right-hand side of the formula in (\ref{formu32}) is called update direction. Let $\boldsymbol{\eta}^{(i)}=(\nabla_{\boldsymbol{\kappa}_{2}}^{H}(\boldsymbol{\Gamma}(\boldsymbol{\kappa}_{2}^{(i)},\boldsymbol{\theta},\mathbf{w}, \mathbf g[t])\mathbf{h}[t]))^{\dagger}(\mathbf{y}[t]-\boldsymbol{\Gamma}(\boldsymbol{\kappa}_{2}^{(i)},\boldsymbol{\theta},\mathbf{w}, \mathbf g[t])\mathbf{h}[t])$. After $\boldsymbol{\eta}^{(i)}$ is given, $\boldsymbol{\kappa}_{2}^{(i+1)}$ is updated as
\begin{align}
\boldsymbol{\kappa}_{2}^{(i+1)}=\boldsymbol{\kappa}_{2}^{(i)}+\lambda^{(i)}\boldsymbol{\eta}^{(i)},\label{formu33}
\end{align}
where $\lambda^{(i)}$ is the step size, which is set according to the Armijo rule in \cite{bertsekas1976goldstein}. Similarly, $\boldsymbol{\kappa}_{1}$ can also be obtained based on the above optimization steps, which will not be repeated in this section.

\section{Convergence Analysis}\label{IV}
In this section, to analyze the convergence of JLBO algorithm, we have the following assumptions on the JLBO algorithm. $\textbf{(1)}$: Gradient matrix $\mathrm{rank}(\nabla \mathbf{g}(\boldsymbol{\kappa};\mathbf{w}))=N$. $\textbf{(2)}$: $\mathbb{E}[\mathbf{n}]=\mathbf{0}$. $\textbf{(3)}$: $N_{S}N_{U}NM\geq 2N(L_{1}+1)+3$ and $N_{S}N_{U}NM\geq 2N(L_{2}+1)+3$. The fisrt assumption means the rank of the gradient matrix
should be equal to the number of unknown parameters, which is usually true for the mmWave system \cite{9053493,fascista2021ris}. The second assumption means the number of pilots should be no less than the number of unknown parameters in the JLBO scheme, which is satisfied by the usual mmWave MIMO systems \cite{9053493,fascista2021ris}.
\begin{theorem}\label{The32}
When assumptions (1)-(3) are satisfied, at each slot $k$, localization error is bounded.
The proof is given in \textbf{Appendix~C}.
\end{theorem}
\textbf{Theorem~2} shows that when the above conditions are satisfied, the error of channel estimation is limited and proportional to the noise power. Thus, channel estimation is convergence.
\begin{theorem}\label{The33}
When assumptions in (1)-(3) are satisifed and the number of iterations is good enough,  the beamforming algorithm converges to a sationary solution.
The proof is given in \textbf{Appendix~D}.
\end{theorem}
To sum up, each sub-algorithm of the proposed JLBO algorithm is convergent, and thus the overall JLBO algorithm is convergent according to [41]-[43].

\section{Numerical Results}\label{V}
 
The simulation results are given in this section to show the positioning performance of the RIS-aided mmWave communication system. The system consists of $5$ BSs,$1$ UE and $1$ RIS. The carrier frequency is {\color{red}$f_{c}=28$~GHz}, the number of transmitting antennas of each BS is $N_{T}=64$, the number of receiving antennas of UE is $N_{U}=4$, and the light speed is $c=3\times 10^{8}$~m/s. In addition, the distances between the adjacent transmitting antenna elements, the RIS elements and the receiving antenna elements are $d_{B}=c/f_{c}/2$~m, $d_{R}=c/f_{c}/2$~m and $d_{U}=c/f_{c}/2$~m, respectively. We assume that UE and $5$ BSs are randomly distributed in a rectangular region of $1\times 1$~$\mathrm{km}^{2}$, and their orientation angles are also random. 
The number of paths from the BS to the RIS is $L_{1}=10$, and the number of paths from the RIS to the UE is $L_{2}=8$. The sampling period is $T_{s}=10$~ns.
{\color{red}Based on\cite{9760391}, the variance of complex channel gains  $\sigma_{g,n}^{2}$ and $\sigma_{h,n}^{2}$ satisfy $\sigma_{g,n}^{2}=10^{-0.5}\nu_{BR}$ and $\sigma_{h,n}^{2}=10^{-0.5\nu_{Rk}}$, where $\nu_{BR}=10^{-6,14-2\log_{10}(d_{BR})}$ and $\nu_{RK}=10^{-6,14-2\log_{10}(d_{Rk})}$, $d_{BR}$ and $d_{Rk}$ represent the distances from the BS to the RIS and the RIS to the $k$-th UE.} {\color{red}In this section, these algorithms are used as baselines:}

{\color{red}\textbf{ML-based algorithm}\cite{brighente2019machine}: This paper proposed a wireless localization algorithm based on machine learning by exploiting support vector machines (SVMs) with typical loss functions. 

\textbf{GD-based algorithm}\cite{keykhosravi2021siso}: This paper proposed a gradient descent (GD) algorithm to design beamforming and the localization problem is solved based on majorize-minimize (MM). 

\textbf{LS-based algorithm}\cite{9593204}: This scheme use a line search (LS) algorithm based on Armijo rule. }
%{}{In addition to the rate, the energy efficiency}, which is defined as the ratio of sum-rate to total consumption power, that is,
%\begin{eqnarray}
%EE=\frac{R}{P+N_{RF}P_{RF}+N_{PS}P_{PS}+N_{r}P_{RIS}},\label{3-68}
%\end{eqnarray}
%where $P_{RF}$ is the power consumption of each RF chian, and $N_{RF}=N{t}$ is an all-digital structure, and when $N_{RF}=N$ is a mixed structure. 
%$P_{PS}$ is the power consumed by each phase shift, and the number of phase shifters is also $N_{PS}$, where it is an all-digital structure when $N_{PS}=0$ and a mixed structure when $N_{PS}=NN_{T}$. $P_{RIS}$ is the power consumed by each reflecting elements. 
%In the simulation, $P_{RF}=300$~mW and $P_{PS}=40$~mW\cite{dai2018hybrid}, $P_{RIS}=10$~mW\cite{huang2019reconfigurable}.

%\begin{figure}[!h]
%\centering
%\includegraphics[scale=0.45]{FIG3_2E.pdf}
%  \vspace{2.0cm}
%  \medskip
  %\caption{Simulated RIS-aided mmWave-NOMA communication scenario in (a). Simulated mmWave-NOMA communication scenario without RIS in (b)\label{fig:3_2}.}
%\end{figure}
\begin{figure}[H]
\centering
\begin{minipage}[t]{0.48\textwidth}
\centering
\includegraphics[scale=0.5]{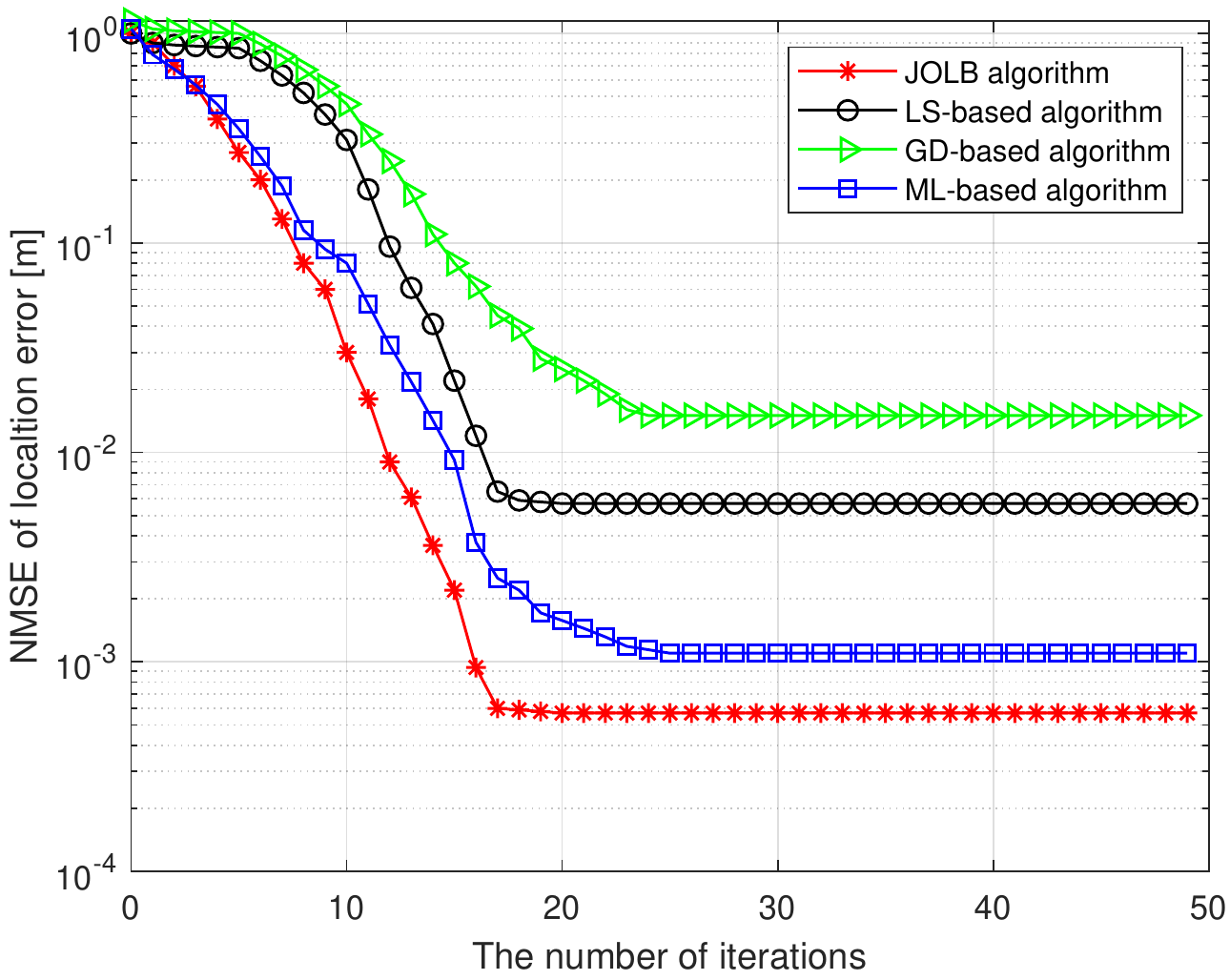}
\caption{NMSE of localization error versus the number of iterations}
\label{fig:v5_1}
\end{minipage}
\begin{minipage}[t]{0.48\textwidth}
\centering
\includegraphics[scale=0.5]{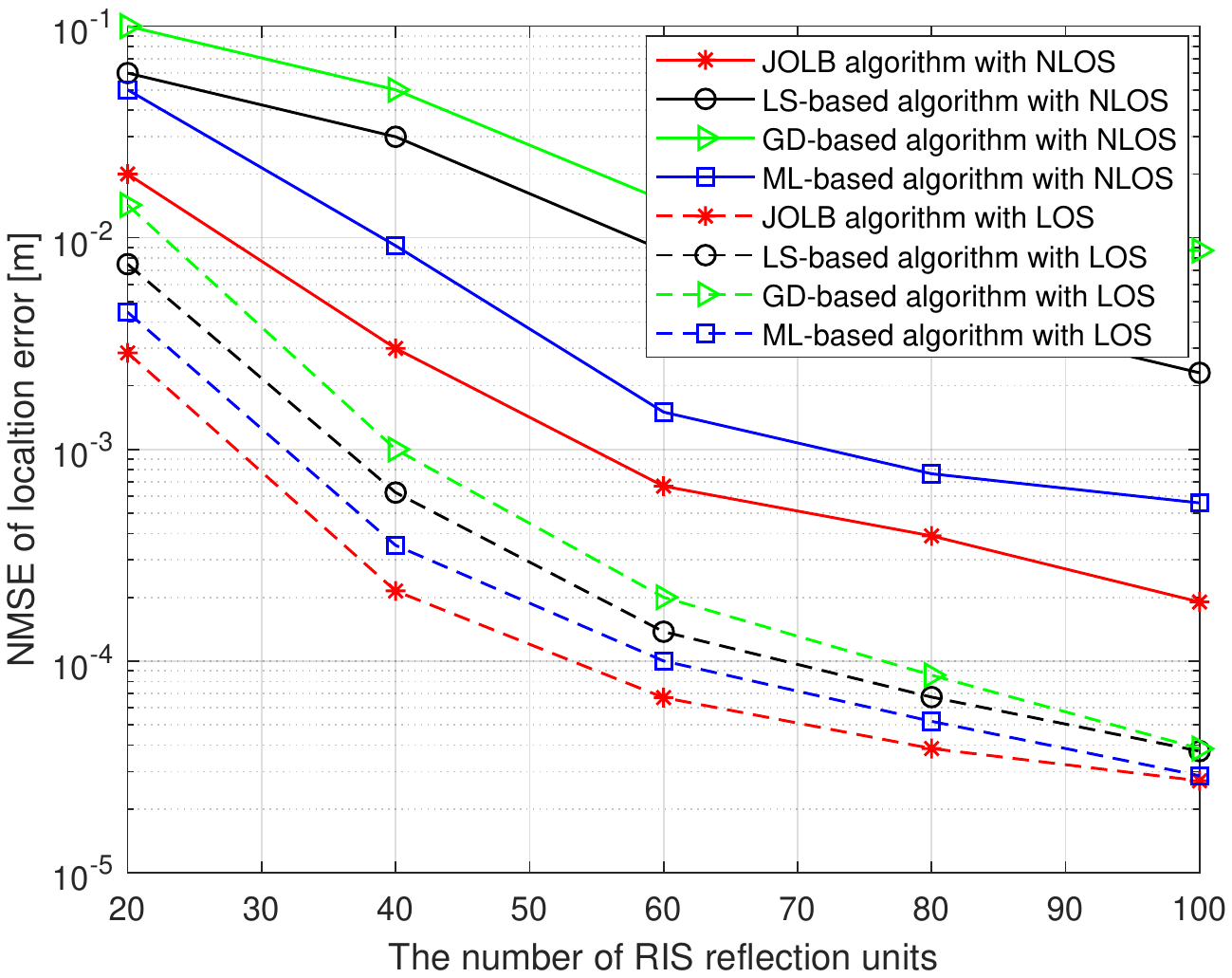}
\caption{NMSE of localization error versus the $N_{R}$}
\label{fig:v5_2}
\end{minipage}
\end{figure}

Fig. \ref{fig:v5_1} presents the normalized mean square error (NMSE) of various localization algorithms on the first slot of a random BF vector. We can see that the NMSE under different iterations decreases monotonically with the increase of iterations. It can be seen that the proposed localization algorithm can achieve a localization error of $0.028$~m. In addition, the proposed CRLB algorithm has a faster convergence rate than existing algorithms, usually achieving a converged NMSE with only several iterations for the CRLB algorithm. This is because the proposed algorithm is an estimation algorithm designed based on a specific parameter estimation problem. At the same time, the fast convergence also reflects the low complexity of the CRLB algorithm, which is attractive for practical applications.

As shown in Fig. \ref{fig:v5_2}, the influence of geometry on positioning and orientation estimation errors is studied as a function of the number of elements in the RIS. The localization error decrease with the increase of the number of RIS reflecting units. In addition, the results also show that the proposed CRLB based beamforming algorithm achieves satisfactory performance improvement in terms of positioning accuracy. It can be observed from the figure that when the channel is a LOS channel, not many RIS reflecting units are needed to achieve high positioning accuracy, this is because the number of LOS is much less than the length of the pilot sequence. This shows that in the actual RIS deployment, the number of RIS reflecting units can be determined according to the channel scenario, which is of great practical value in the actual deployment. Finally, with the increase of the number of channel paths, the performance of all algorithms will decline, but the performance loss of the JLBO algorithm proposed in this paper is the least.

%In addition, the RIS-aided scheme can greatly improve energy efficiency, as shown in Fig.~\ref{fig:3_8}. It can also be found that the energy efficiency of all-digital mmWave-NOMA and hybrid mmWave-NOMA without RIS is lower. In particular, when the minimum rate constraint does not exceed $1bps/Hz$, the energy efficiency of the proposed RIS-aided mmWave-NOMA scheme can reach nearly $0.6$ times that of the mmWave-NOMA without the RIS. The reason for this phenomenon is that RIS is an aided device with less energy consumption, so the RIS does not need to consume more energy to improve beam gain, which further illustrates the advantages of RIS in improving the performance of communication systems.

%\begin{figure}[!h]
%\centering
%\includegraphics[scale=0.5]{FIG3_9E.pdf}
%\caption{Sum-rate versus total power,$\gamma=1$~bps/Hz.\vspace{-15pt}}
%\label{fig:3_9}
%\end{figure}
%\begin{figure}
%\centering
%\includegraphics[scale=0.5]{FIG3_10E.pdf}
%\caption{Energy efficiency versus total power, $\gamma=1$~bps/Hz.\vspace{-15pt}}
%\label{fig:3_10}
%\end{figure}\vspace{-15pt}

\begin{figure}[H]
\centering
\begin{minipage}[t]{0.48\textwidth}
\centering
\includegraphics[scale=0.5]{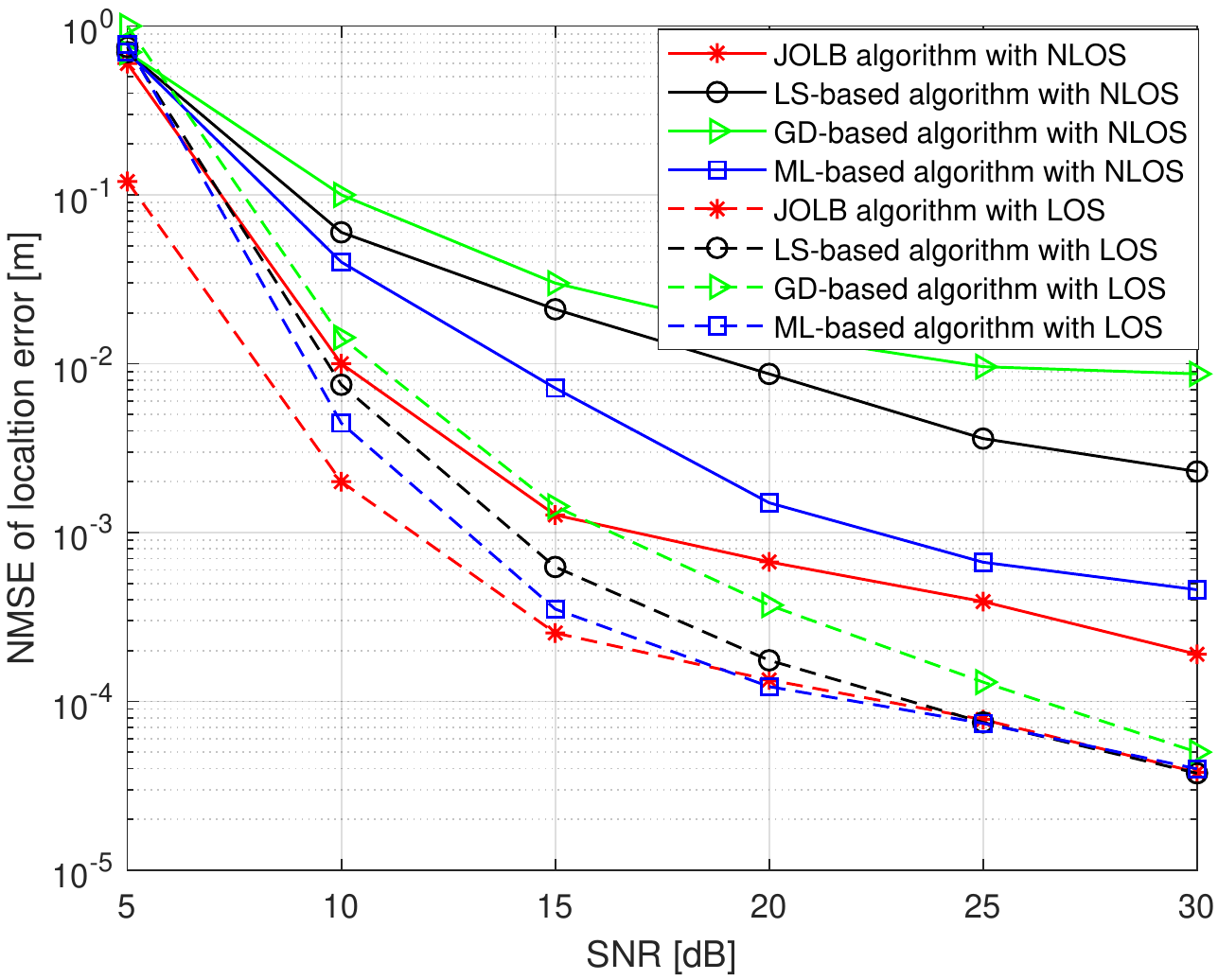}
\caption{NMSE of localization error versus SNR}
\label{fig:v5_3}
\end{minipage}
\begin{minipage}[t]{0.48\textwidth}
\centering
\includegraphics[scale=0.55]{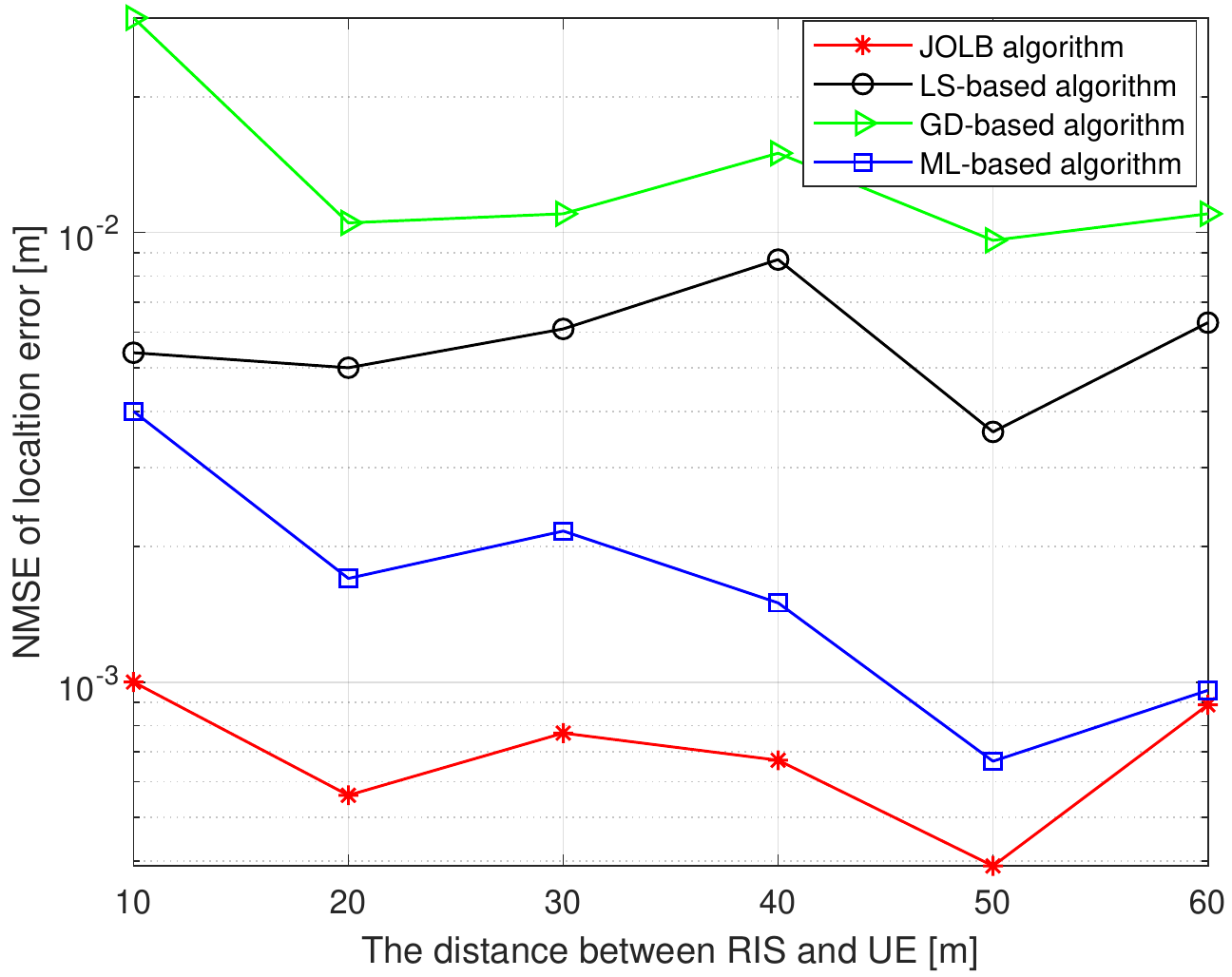}
\caption{NMSE of localization error versus the distance $d_{AR}$}
\label{fig:v5_4}
\end{minipage}
\end{figure}

In Fig. \ref{fig:v5_3}, we report the NMSE of the MS position estimation calculated in 200 independent Monte Carlo experiments as a function of SNR. The proposed JLBO algorithm can achieve high estimation accuracy when the SNR is $15$~dB, and the NMSE tends to decrease with the further increase of the SNR, but because of the inherent suboptimality of JLBO, it can not strictly reach perfect localization error. In addition, once the JLBO algorithm gets the desired initialization value, the NMSE of the joint maximum likelihood estimator will drop immediately, thus achieving high positioning accuracy with $10$~dB SNR. In addition, it is not difficult to find from the figure that when the channel of the system is LOS channel, the performance of the proposed JLBO algorithm is still better than the existing algorithms in the low SNR scenario, because compared with the existing algorithms, CRLB is used to optimize the beam in the beamforming phase. In the high SNR scenario, the performance of the proposed JLBO algorithm is consistent with that of the existing algorithms, because the number of variables in the LOS channel is much smaller than the pilot training length, so the existing algorithms can also achieve high positioning accuracy.

In Fig. \ref{fig:v5_4}, We give the relationship between the positioning accuracy and the distance between the RIS and the BS. It can be seen from the figure that compared with the existing localization algorithms, the estimation accuracy of JLBO algorithm is higher. In addition, because we combine the position information to update the beam, the positioning accuracy is constantly modified. 
This further reduces the reduction of estimation accuracy caused by inaccurate acquisition of RIS position information. 
It can be seen that the JLBO algorithm is more accurate than the position estimated by random beamforming. 
This is because the position information will be updated every time the beam information is updated, and the position calculation based on the geometric relationship will be more accurate. Finally, we find that the relationship between NMSE and the distance between the RIS and the BS is not linear. 
When the distance between the RIS and the BS is close, it is shown that NMSE decreases. 
When the distance between the RIS and the BS is more than 20 m, the NMSE of positioning error will increase. 
This is because the path calculation decreases as the distance increases, however, when the distance between the RIS and the BS reaches 30 m, the estimation accuracy continues to decrease, because although the path loss between the BS and the RIS increases, the path loss between the RIS and the UE decreases and is greater than that between the BS and the RIS. 
This result shows that the positioning accucary is strongly depended on the deployment of the RIS., and the actual RIS deployment provides theoretical support.

\begin{figure}[H]
  \centering
  % include first image
  \includegraphics[scale=0.5]{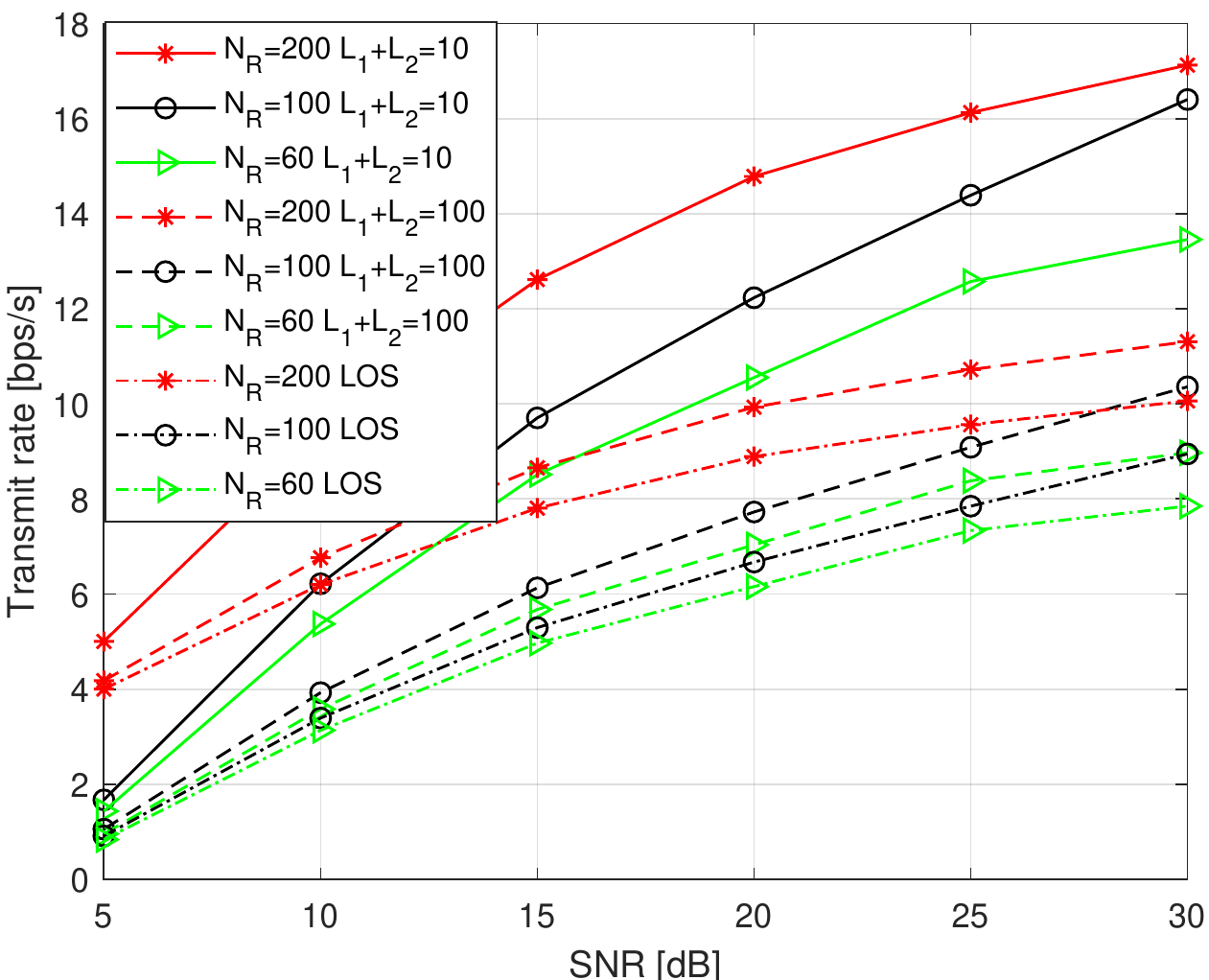}
  \caption{System rate versus SNR with different $N_{R}$}
\label{fig:v5_5}
\end{figure}
In Fig.~\ref{fig:v5_5}, we find that the system rate grows with the increase of SNR, which is because in the high SNR scenario, the positioning error decreases and the beam direction is aligned with the user. 
It can also be observed from the figure that the system rate rises with the increase of the number of RIS reflecting units, because the beam and the user can be better aligned with more RIS reflecting units, so that more beam energy converges to the user. 
It is noteworthy that when the NLOS path reduces while other parameters remain the same, the performance improvement is more significant because our training length is long enough. However, when the number of NLOS exceeds the pilot length, the transmission rate performance of the system is even worse than that of the LOS channel, because the number of positioning parameters is much larger than the pilot length, which leads to the decline of positioning accuracy, and further leads to the performance degradation caused by beam pointing and user misalignment.

\section{Conclusion and Future Works}\label{VI}
In this paper, we studied the successive localization and beamforming design of the RIS-aided mmWave communication system. Due to the cascaded channel in the system, the problem was a multivariable coupled non-convex problem. Therefore, this paper presented a new alternating optimization algorithm to solve this problem. Firstly, the channel gain of the cascaded channel was estimated, and then the position parameters were estimated. Finally, transmit beamforming and RIS reflecting beamforming were designed based on the CRLB. In addition, the convergence of the proposed scheme was proved. Simulation results showed that the performance of the proposed JLBO scheme can be greatly improved compared with the existing joint localization and beamforming methods.

\appendices
\section{Proof of \textbf{Theorem~1}}\label{appA}
According to the formulas in (\ref{formu1}), (\ref{formu2}) and (\ref{formu13}), the received signals from $N$ BSs at $N_{S}$ subcarriers are rewritten.
\begin{align}
\left[
 \begin{matrix}
\mathbf{y}_{1}[t,1,m]\\
\vdots\\
\mathbf{y}_{N}[t,N_{S},m]
  \end{matrix}
  \right]=\left[
 \begin{matrix}\mathbf{A}_{U,1}[1]&\ldots&\mathbf{0}\\
  \vdots&\ddots&\vdots\\
  \mathbf{0}&\ldots&\mathbf{A}_{U,N}[N_{S}]  \end{matrix}
  \right]\left[
 \begin{matrix}
\tilde{\mathbf{H}}_{1}[t,1]&\ldots&\mathbf{0}\\
\vdots&\ddots&\vdots\\
\mathbf{0}&\ldots&\tilde{\mathbf{H}}_{N}[t,N_{S}]
  \end{matrix}
  \right] 
\nonumber\\
\left[
 \begin{matrix}\mathbf{A}_{\bar{R},1}^{H}[1]&\ldots&\mathbf{0}\\
  \vdots&\ddots&\vdots\\
  \mathbf{0}&\ldots&\mathbf{A}_{\bar{R},N}^{H}[N_{S}]  \end{matrix}
  \right]\left[\begin{matrix}
\boldsymbol{\Theta}\mathbf{G}_{1}[t,1]\mathbf{w}_{1}[1,m]\\
\vdots\\
\boldsymbol{\Theta}\mathbf{G}_{N}[t,N_{S}]\mathbf{w}_{N}[N_{S},m]
  \end{matrix}
  \right]+\left[\begin{matrix}
\mathbf{n}_{1}[t,1,m]\\
\vdots\\
\mathbf{n}_{N}[t,N_{S},m]
  \end{matrix}
  \right].\label{formuA1}%
\end{align}

For brevity, let $
\mathbf{b}_{n}[t,j,m]\in\mathbb{C}^{N_{R}\times N_{T}}=\boldsymbol{\Theta}\mathbf{G}_{n}[t,j]\mathbf{w}_{n}[j,m]$, (\ref{formuA1}) is rewritten as
\begin{align}
\left[
 \begin{matrix}
\mathbf{y}_{1}[t,1,m]\\
\vdots\\
\mathbf{y}_{N}[t,N_{S},m]
  \end{matrix}
  \right]=\left[
 \begin{matrix}\mathbf{A}_{U,1}[1]&\ldots&\mathbf{0}\\
  \vdots&\ddots&\vdots\\
  \mathbf{0}&\ldots&\mathbf{A}_{U,N}[N_{S}]  \end{matrix}
  \right]\left[
 \begin{matrix}
\tilde{\mathbf{H}}_{1}[t,1]&\ldots&\mathbf{0}\\
\vdots&\vdots&\vdots\\
\mathbf{0}&\ldots&\tilde{\mathbf{H}}_{N}[t,N_{S}]
  \end{matrix}
  \right] 
\nonumber\\
\left[
 \begin{matrix}\mathbf{A}_{\bar{R},1}^{H}[1]&\ldots&\mathbf{0}\\
  \vdots&\ddots&\vdots\\
  \mathbf{0}&\ldots&\mathbf{A}_{\bar{R},N}^{H}[N_{S}]  \end{matrix}
  \right]\left[\begin{matrix}
\mathbf{b}_{1}[t,1,m]\\
\vdots\\
\mathbf{b}_{N}[t,N_{S},m]
  \end{matrix}
  \right]+\left[\begin{matrix}
\mathbf{n}_{1}[t,1,m]\\
\vdots\\
\mathbf{n}_{N}[t,N_{S},m]
  \end{matrix}
  \right].\label{formuA2}
\end{align}
According to (\ref{formu6}) and (\ref{formu8}), it is rewritten as
\begin{align}
(\ref{formuA2})=\sqrt{N_{R}N_{U}}\left[\begin{matrix}
\mathbf{a}_{U,1}(\theta_{U,0,1})&\ldots&\mathbf{a}_{U,1}(\theta_{U,L_{2},1})&\mathbf{0}&\ldots&\mathbf{0}\\
\vdots&\vdots&\vdots&\vdots&\vdots&\vdots\\
\mathbf{0}&\ldots&\mathbf{0}&\mathbf{a}_{U,N}(\theta_{U,0,N_{S}})&\ldots&\mathbf{a}_{U,N}(\theta_{U,L_{2},N_{S}})
  \end{matrix}
  \right]\nonumber\\
\left[\begin{matrix}
e^{-\bar{j}2\pi\frac{1}{N_{S}T_{S}}\tau_{0,1}}&\ldots&\mathbf{0}\\
\vdots&\ddots&\vdots\\
\mathbf{0}&\ldots&e^{-\bar{j}2\pi\frac{N_{S}}{N_{S}T_{S}}\tau_{L_{2},N_{S}}}
  \end{matrix}
  \right]\left[\begin{matrix}
h_{0,1}[t]&\ldots&\mathbf{0}\\
\vdots&\ddots&\vdots\\
\mathbf{0}&\ldots&h_{L_{2},N}[t]
  \end{matrix}
  \right]\nonumber\\
\left[\begin{matrix}
\mathbf{a}_{\bar{R},1}^{H}(\theta_{\bar{R},0,1})&\ldots&\mathbf{0}\\
\vdots&\vdots&\vdots\\
\mathbf{a}_{\bar{R},1}^{H}(\theta_{\bar{R},L_{2},1})&\ldots&\mathbf{0}\\
\vdots&\vdots&\vdots\\
\mathbf{0}&\ldots&\mathbf{a}_{\bar{R},N_{S}}^{H}(\theta_{\bar{R},0,N})\\
\vdots&\vdots&\vdots\\
\mathbf{0}&\ldots&\mathbf{a}_{\bar{R},N_{S}}^{H}(\theta_{\bar{R},L_{2},N})
  \end{matrix}
  \right]\left[\begin{matrix}
\mathbf{b}_{1}[t,1,m]\\
\vdots\\
\mathbf{b}_{N}[t,N_{S},m]
  \end{matrix}
  \right]+\left[\begin{matrix}
\mathbf{n}_{1}[t,1,m]\\
\vdots\\
\mathbf{n}_{N}[t,N_{S},m]
  \end{matrix}
  \right].\label{formuA3}
\end{align}
In order to linearize the expression in (\ref{formuA3}), we use vectorization operation to rewrite (\ref{formuA3}).
\begin{align}
\mathrm{vec}((\ref{formuA3}))=\sqrt{N_{R}N_{U}}\left(\left[\begin{matrix}
\mathbf{a}_{\bar{R},1}^{H}(\theta_{\bar{R},0,1})\mathbf{b}_{1}[t,1,m]\\
\vdots\\
\mathbf{a}_{\bar{R},1}^{H}(\theta_{\bar{R},L_{2},1})\mathbf{b}_{1}[t,1,m]\\
\vdots\\
\mathbf{a}_{\bar{R},N_{S}}^{H}(\theta_{\bar{R},0,N})\mathbf{b}_{N}[t,N_{S},m]\\
\vdots\\
\mathbf{a}_{\bar{R},N_{S}}^{H}(\theta_{\bar{R},L_{2},N})\mathbf{b}_{N}[t,N_{S},m]
  \end{matrix}
  \right]^{T}\right.\otimes~~~~~~~~~~~~~~~~~~~~~~~~~~~~~~~~\nonumber\\
\left(\left[\begin{matrix}
\mathbf{a}_{U,1}(\theta_{U,0,1})&\ldots&\mathbf{a}_{U,1}(\theta_{U,L_{1},1})&\mathbf{0}&\ldots&\mathbf{0}\\
\vdots&\vdots&\vdots&\vdots&\vdots&\vdots\\
\mathbf{0}&\ldots&\mathbf{0}&\mathbf{a}_{U,N}(\theta_{U,0,N_{c}})&\ldots&\mathbf{a}_{U,N}(\theta_{U,L_{1},N_{c}})
  \end{matrix}
  \right]\right.~~~~~~~~~~~~~~~~~~~~~~~~~~~~~~~~~~~~~~~~~~~~~~\nonumber\\
  \left.\left.\left[\begin{matrix}
e^{-\bar{j}2\pi\frac{1}{N_{S}T_{S}}\tau_{0,1}}&\ldots&\mathbf{0}\\
\vdots&\ddots&\vdots\\
\mathbf{0}&\ldots&e^{-\bar{j}2\pi\frac{N_{S}}{N_{S}T_{S}}\tau_{L_{2},N_{S}}}
  \end{matrix}
  \right]\right)\right)\mathrm{vec}\left(\left[\begin{matrix}
h_{0,1}[k]&\ldots&\mathbf{0}\\
\vdots&\ddots&\vdots\\
\mathbf{0}&\ldots&h_{L_{2},N}[k]
  \end{matrix}
  \right]\right) +\left[\begin{matrix}
\mathbf{n}_{1}[t,1,1]\\
\vdots\\
\mathbf{n}_{N}[t,N_{S},m]
  \end{matrix}
  \right].~~~~~~~~~~~~~~~~~~~~~~~\label{formuA4}
  \end{align}
The matrix element of the first Kronecker product operation on the right-hand side of the formula (\ref{formuA4}) is expressed as
\begin{align}
&A_{i_{1},i_{2}}[m]=\sum_{\bar{t}=1}^{N_{R}} b_{n}[t,j,m]^{(\bar{t})}e^{-\bar{j}2\pi\frac{j}{N_{S}T}}a_{R,n}^{*(\bar{t})}(\theta_{R,l_{2},j})a_{U,n}^{(r)}(\theta_{U,l_{2},j}),\nonumber\\
&1\leq i_{1}\leq N_{S}N_{U}N,
1\leq i_{2}\leq (N_{S}N(L_{2}+1))^{2},\label{formuA5}
\end{align}
where $\bar{t}\in\{1,\ldots,N_{R}\}$, the relationship between $i_{1}$,$i_{2}$ and $j$,$n$,$l_{2}$,$r$ are expressed as
\begin{align}
&j=\left\{
 \begin{array}{lr}\lfloor i_{1}/N_{U}N\rfloor+1, & i_{1}\neq N_{U}N,2N_{U}N\ldots,N_{S}N_{U}N,\\
i_{1}/N_{U}N+1, & i_{1}=N_{U}N,2N_{U}N\ldots,N_{S}N_{U}N,  
\end{array}
\right.\nonumber\\
&n=\left\{
 \begin{array}{lr}\lfloor i_{1}/N_{U}N_{S}\rfloor+1, & i_{1}\neq N_{U}N_{S},2N_{U}N_{S}\ldots,N_{S}N_{U}N,\\
i_{1}/N_{U}N_{S}+1, & i_{1}=N_{U}N_{S},2N_{U}N_{S}\ldots,N_{S}N_{U}N,  
\end{array}
\right.\nonumber\\
&l_{2}=\left\{
 \begin{array}{lr}\lfloor i_{2}/(NN_{S})^{2}/(L_{2}+1)\rfloor+1, & i_{2}\neq (NN_{S})^{2}(L_{2}+1),2(NN_{S})^{2}(L_{2}+1)\ldots,(NN_{S})^{2}(L_{2}+1)^{2},\\
i_{2}/(NN_{S})^{2}/(L_{2}+1), & i_{2}= (NN_{S})^{2}(L_{2}+1),2(NN_{S})^{2}(L_{2}+1)\ldots,(NN_{S})^{2}(L_{2}+1)^{2},
\end{array}
\right.\nonumber\\
&r=\left\{
 \begin{array}{lr}\lfloor i_{1}/N_{S}N\rfloor+1, & i_{1}\neq N_{S}N,2N_{S}N\ldots,N_{S}N_{U}N,\\
i_{1}/N_{S}N+1, & i_{1}=N_{S}N,2N_{S}N\ldots,N_{S}N_{U}N,  
\end{array}
\right.\label{formuA_4}
\end{align}
where $b_{n}[t,j,m]^{(i)}$ is the $i$-th element of vector $\mathbf{b}_{n}[t,j,m]$. $a_{R,j}^{*(i)}(\theta_{R,l_{2},n})$ and $a_{U,j}^{(1)}(\theta_{U,l_{2},n})$ are the $i$-th element of vector $\mathbf{a}_{R,j}(\theta_{R,l_{2},n})$ and vector $\mathbf{a}_{U,j}(\theta_{U,l_{2}-1,n})$, respectively. Further simplify
$\mathbf{y}_{1}[t,1,1],\ldots,\mathbf{y}_{N}[t,N_{S},M]$, we have
 \begin{align}
&\mathbf{\Gamma}(\boldsymbol{\kappa},\boldsymbol{\theta},\mathbf{w},\mathbf{g}[t])\in\mathbb{C}^{N_{U}N_{S}NM\times N(L_{2}+1)}=\left[\begin{matrix}
\boldsymbol{\Gamma}_{1}&\ldots&\mathbf{0}\\
\vdots&\ddots&\vdots\\
\mathbf{0}&\ldots&\boldsymbol{\Gamma}_{N}
  \end{matrix}
  \right],\label{formuA_6}\\
&\boldsymbol{\Gamma}_{n}\in\mathbb{C}^{N_{U}N_{S}M\times(L_{2}+1)}=[\boldsymbol{\gamma}_{n}^{(1)}[1,1],\ldots,\boldsymbol{\gamma}_{n}^{(N_{U})}[N_{S},M]]^{T},\label{formuA6}\\
&\boldsymbol{\gamma}_{n}^{(r)}[j,m]\in\mathbb{C}^{(L_{2}+1)\times 1}=[\gamma_{0,n}^{(r)}[j,m],\ldots,\gamma_{L_{2},n}^{(r)}[j,m]]^{T},\gamma_{l_{2},n}^{(r)}[j,m]=\mathbf{b}_{n}^{H}[t,j,m]\boldsymbol{\mu}_{l_{2},j,n}^{(r)},\label{formuA7}\\
&\mathbf{b}_{n}[t,j,m]\in\mathbb{C}^{N_{R}\times 1}=\boldsymbol{\Theta}\mathbf{G}_{n}[t,j]\mathbf{w}_{n}[j,m],\boldsymbol{\mu}_{l_{2},j,n}^{(r)}\in\mathbb{C}^{N_{T}\times 1}=[\mu_{l_{2},j,n}^{(r,1)},\ldots,\mu_{l_{2},j,n}^{(r,N_{R})}]^{T}, \label{formuA8}\\
&\mu_{l_{2},j,n}^{(r,\bar{t})}=\sqrt{N_{R}N_{U}}e^{-\bar{j}2\pi\frac{j}{N_{S}T_{S}}\tau_{l_{2},n}}a_{R,n}^{*(\bar{t})}(\theta_{R,l_{2},j})a_{U,n}^{(r)}(\theta_{U,l_{2},j}),
 \mathbf{h}[t]\in\mathbb{C}^{N(L_{2}+1)\times 1}=[h_{1,0}[t],...,h_{N,L_{2}}[t]]^{T}.\label{formuA9}
 \end{align}

\section{Proof of \textbf{Theorem~\ref{The31}}}\label{appD}
According to estimation theory in \cite{kay1993fundamentals}, the FIMs of problem in (\ref{formu34}) and (\ref{formu35}) are expressed as
\begin{align}
\boldsymbol{\mathcal{J}}(\boldsymbol{\kappa}_{1},\mathbf{h}[t];\mathbf{w},\boldsymbol{\theta},\mathbf{g}[t])\in\mathbb{C}^{(2N(L_{2}+1)+3+N(L_{2}+1))\times(2N(L_{2}+1)+3+N(L_{2}+1))}=\sigma^{-2}\sum_{r,j,m}~~~~~~~~~~~~~~~~~~~~~~~\nonumber\\
\left[
 \begin{matrix}
   \boldsymbol{\Xi}_{j,m}^{(r)}(\boldsymbol{\kappa}_{2};\mathbf{w},\boldsymbol{\theta},\mathbf{g}[t])\mathbf{h}^{*}[t]\mathbf{h}^{T}[t](\boldsymbol{\Xi}_{j,m}^{(r)}(\boldsymbol{\kappa};\mathbf{w},\boldsymbol{\theta},\mathbf{g}[t]))^{H} & \boldsymbol{\Xi}_{j,m}^{(r)}(\boldsymbol{\kappa};\mathbf{w},\boldsymbol{\theta},\mathbf{g}[t])\mathbf{M}^{H}[t]\boldsymbol{\Gamma}_{j,m}^{(r)}(\boldsymbol{\kappa};\mathbf{w},\boldsymbol{\theta},\mathbf{g}[t]) \\
   (\boldsymbol{\Gamma}_{j,m}^{(r)}(\boldsymbol{\kappa};\mathbf{w},\boldsymbol{\theta},\mathbf{g}[t]))^{H}\mathbf{M}[t](\boldsymbol{\Xi}_{j,m}^{(r)}(\boldsymbol{\kappa};\mathbf{w},\boldsymbol{\theta},\mathbf{g}[t]))^{H} & (\boldsymbol{\Gamma}_{j,m}^{(r)}(\boldsymbol{\kappa};\mathbf{w},\boldsymbol{\theta},\mathbf{g}[t]))^{H}\boldsymbol{\Gamma}_{j,m}^{(r)}(\boldsymbol{\kappa};\mathbf{w},\boldsymbol{\theta},\mathbf{g}[t]) 
  \end{matrix}
  \right] \label{formuA10}
\end{align}
and
\begin{align}
\boldsymbol{\mathcal{I}}(\boldsymbol{\kappa},\mathbf{g}[t];\mathbf{w},\boldsymbol{\theta},\mathbf{h}[t])\in\mathbb{C}^{(2N(L_{1}+1)+3+N(L_{1}+1))\times(2N(L_{1}+1)+3+N(L_{1}+1))}=\sigma^{-2}\sum_{r,j,m}~~~~~~~~~~~~~~~~~~~~~~~~~~\nonumber\\
\left[
 \begin{matrix}
   \boldsymbol{\Psi}_{j,m}^{(r)}(\boldsymbol{\kappa};\mathbf{w},\boldsymbol{\theta},\mathbf{h}[t])\mathbf{g}^{*}[t]\mathbf{g}^{T}[t](\boldsymbol{\Psi}_{j,m}^{(r)}(\boldsymbol{\kappa};\mathbf{w},\boldsymbol{\theta},\mathbf{h}[t]))^{H} & \boldsymbol{\Psi}_{j,m}^{(r)}(\boldsymbol{\kappa};\mathbf{w},\boldsymbol{\theta},\mathbf{h}[t])\mathbf{S}^{H}[t]\boldsymbol{\Lambda}_{j,m}^{(r)}(\boldsymbol{\kappa};\mathbf{w},\boldsymbol{\theta},\mathbf{h}[t]) \\
   (\boldsymbol{\Lambda}_{j,m}^{(r)}(\boldsymbol{\kappa};\mathbf{w},\boldsymbol{\theta},\mathbf{h}[t]))^{H}\mathbf{S}[t](\mathbf{K}_{j,m}^{(r)}(\boldsymbol{\kappa};\mathbf{w},\boldsymbol{\theta},\mathbf{h}[t]))^{H} & (\boldsymbol{\Lambda}_{j,m}^{(r)}(\boldsymbol{\kappa};\mathbf{w},\boldsymbol{\theta},\mathbf{h}[t]))^{H}\boldsymbol{\Gamma}_{j,m}^{(r)}(\boldsymbol{\kappa};\mathbf{w},\boldsymbol{\theta},\mathbf{h}[t])
  \end{matrix}
  \right], \label{formuA11}
\end{align}
where $\mathbf{M}[t]\in\mathbb{C}^{N\times N(L_{2}+1)}=\mathbf{1}_{N}\otimes\mathbf{h}^{T}[t]$, $\mathbf{S}[t]\in\mathbb{C}^{N\times N(L_{1}+1)}=\mathbf{1}_{N}\otimes\mathbf{g}^{T}[t]$. $\boldsymbol{\Gamma}_{j,m}^{(r)}(\boldsymbol{\kappa}_{2};\mathbf{w},\boldsymbol{\theta},\mathbf{g}[t])=[\boldsymbol{\gamma}_{n}^{(r)}[j,m],\ldots,\boldsymbol{\gamma}_{n}^{(r)}[j,m]]^{T}$ and $\boldsymbol{\Lambda}_{j,m}^{(r)}(\boldsymbol{\kappa}_{1};\mathbf{w},\boldsymbol{\theta},\mathbf{h}[t]))[\bar{\boldsymbol{\gamma}}_{n}^{(r)}[j,m],\ldots,$ $\bar{\boldsymbol{\gamma}}_{n}^{(r)}[j,m]]^{T}$.
According to expressions in (\ref{formuA10}) and (\ref{formuA11}),  $\mathbb{E}_{\mathbf{h}[t]}[\boldsymbol{\mathcal{J}}(\boldsymbol{\kappa}_{2},\mathbf{h}[t];\mathbf{w},\boldsymbol{\theta},\mathbf{g}[t])^{-1}]$ and $\mathbb{E}_{\mathbf{g}[t]}[\boldsymbol{\mathcal{I}}(\boldsymbol{\kappa}_{1}$ $,\mathbf{g}[t];\mathbf{w},\boldsymbol{\theta},\mathbf{h}[t])^{-1}]$ satisfy the Jensen's inequality, therefore, we have
\begin{align}
\mathbb{E}_{\mathbf{h}[t]}[\boldsymbol{\mathcal{J}}(\boldsymbol{\kappa}_{2},\mathbf{h}[t];\mathbf{w},\boldsymbol{\theta},\mathbf{g}[t])^{-1}]\succeq\sigma^{2}\Bigg(\sum_{r,j,m}~~~~~~~~~~~~~~~~~~~~~~~~~~~~~~~~~~~~~~~~~~~~~~~~~~~~~~~~~~~~~~\nonumber\\
\left[
 \begin{matrix}
   \boldsymbol{\Xi}_{j,m}^{(r)}(\boldsymbol{\kappa}_{2};\mathbf{w},\boldsymbol{\theta},\mathbf{g}[t])\boldsymbol{\Sigma}(\boldsymbol{\Xi}_{j,m}^{(r)}(\boldsymbol{\kappa}_{2};\mathbf{w},\boldsymbol{\theta},\mathbf{g}[t]))^{H} & \mathbf{0}\\
   \mathbf{0}& (\boldsymbol{\Gamma}_{j,m}^{(r)}(\boldsymbol{\kappa}_{2};\mathbf{w},\boldsymbol{\theta},\mathbf{g}[t]))^{H}\boldsymbol{\Gamma}_{j,m}^{(r)}(\boldsymbol{\kappa}_{2};\mathbf{w},\boldsymbol{\theta},\mathbf{g}[t]) 
  \end{matrix}
  \right]\Bigg)^{-1} \nonumber
\end{align}
and
\begin{align}
\mathbb{E}_{\mathbf{g}[t]}[\boldsymbol{\mathcal{I}}(\boldsymbol{\kappa}_{1},\mathbf{g}[t];\mathbf{w},\boldsymbol{\theta},\mathbf{h}[t])^{-1}]\succeq\sigma^{2}\Bigg(\sum_{r,j,m}~~~~~~~~~~~~~~~~~~~~~~~~~~~~~~~~~~~~~~~~~~~~~~~~~~~~~~~~~~~~~~~~~\nonumber\\
\left[
 \begin{matrix}
   \boldsymbol{\Psi}_{j,m}^{(r)}(\boldsymbol{\kappa}_{1};\mathbf{w},\boldsymbol{\theta},\mathbf{h}[t])\boldsymbol{\Pi}(\boldsymbol{\Psi}_{j,m}^{(r)}(\boldsymbol{\kappa}_{1};\mathbf{w},\boldsymbol{\theta},\mathbf{h}[t]))^{H} & \mathbf{0}\\
   \mathbf{0}& (\boldsymbol{\Lambda}_{j,m}^{(r)}(\boldsymbol{\kappa}_{1};\mathbf{w},\boldsymbol{\theta},\mathbf{h}[t]))^{H}\boldsymbol{\Lambda}_{j,m}^{(r)}(\boldsymbol{\kappa}_{1};\mathbf{w},\boldsymbol{\theta},\mathbf{h}[t])
  \end{matrix}
  \right]\Bigg)^{-1}, \label{formuA13}
\end{align}
where $\boldsymbol{\Sigma}\in\mathbb{C}^{N(L_{2}+1)\times N(L_{2}+1)}=\mathbb{E}[\mathbf{h}^{*}[t]\mathbf{h}^{T}[t]]$ and $\boldsymbol{\Pi}\in\mathbb{C}^{N(L_{1}+1)\times N(L_{1}+1)}=\mathbb{E}[\mathbf{g}^{*}[t]\mathbf{g}^{T}[t]]$
\begin{align}
\boldsymbol{\Xi}_{j,m}^{(r)}(\boldsymbol{\kappa}_{2};\mathbf{w},\boldsymbol{\theta},\mathbf{g}[t])\in\mathbb{C}^{(2N(L_{2}+1)+3)\times(N(L_{2}+1))}=\left[
 \begin{matrix}
\boldsymbol{\xi}_{l_{2},n}^{(r)}[j,m]|\forall~l_{2}=0,\ldots,L_{2},\forall~n=1,\ldots,N
  \end{matrix}
  \right],\nonumber\\
\boldsymbol{\Psi}_{j,m}^{(r)}(\boldsymbol{\kappa}_{1};\mathbf{w},\boldsymbol{\theta},\mathbf{h}[t])\in\mathbb{C}^{(2N(L_{1}+1)+3)\times(2N(L_{1}+1))}=\left[
 \begin{matrix}
\boldsymbol{\nu}_{l_{1},n}^{(r)}[j,m]|\forall~l_{1}=0,\ldots,L_{1},\forall~n=1,\ldots,N
  \end{matrix}
  \right].\label{formuA15}
\end{align}
$\boldsymbol{\xi}_{l_{2},n}^{(r)}[j,m]\in\mathbb{C}^{(2N(L_{2}+1)+3))\times 1}$ and $\boldsymbol{\nu}_{l_{1},n}^{(r)}[j,m]\in\mathbb{C}^{(2N(L_{1}+1)+3))\times 1}$ are denoted as
\begin{align}
\boldsymbol{\xi}_{l_{2},n}^{(r)}[j,m]=\left[
 \begin{matrix}
\sum_{\bar{t}=1}^{N_{R}}b_{n}^{(\bar{t})}[j,m]\mu_{l_{2},j,n}^{(r,\bar{t})*}\mathbf{p}_{l_{2},j,n}^{(r,\bar{t})}\\
\sum_{\bar{t}=1}^{N_{R}}b_{n}^{(\bar{t})}[j,m]\mu_{l_{2},j,n}^{(r,\bar{t})*}o_{l_{2},j,n}^{(r,\bar{t})}\\
\sum_{\bar{t}=1}^{N_{R}}b_{n}^{(\bar{t})}[j,m]\mu_{l_{2},j,n}^{(r,\bar{t})*}\mathbf{q}_{l_{2},j,n}^{(r,\bar{t})}
  \end{matrix}
  \right],\label{formuA16}
\end{align}
\begin{align}
\boldsymbol{\nu}_{l_{1},n}^{(r)}[j,m]=\left[
 \begin{matrix}
\sum_{\tilde{t}=1}^{N_{T}}w_{n}^{(\tilde{t})}[t,j,m]\bar{\mu}_{l_{1},j,n}^{(r,\tilde{t})*}\sum_{\bar{t}=1}^{N_{R}}b_{n}[t,j]^{(r,\bar{t})}a_{R,j}^{(\bar{t})}(\theta_{R,l_{1},n})\bar{\mathbf{p}}_{l_{1},j,n}^{(\tilde{t},\bar{t})}\\
\sum_{\tilde{t}=1}^{N_{T}}w_{n}^{(\tilde{t})}[t,j,m]\bar{\mu}_{l_{1},j,n}^{(r,\tilde{t})*}\sum_{\bar{t}=1}^{N_{R}}b_{n}[t,j]^{(r,\bar{t})}a_{R,j}^{(\bar{t})}(\theta_{R,l_{1},n})\bar{o}_{l_{1},j,n}^{(\tilde{t},\bar{t})}\\
\sum_{\tilde{t}=1}^{N_{T}}w_{n}^{(\tilde{t})}[t,j,m]\bar{\mu}_{l_{1},j,n}^{(r,\tilde{t})*}\sum_{\bar{t}=1}^{N_{R}}b_{n}[t,j]^{(r,\bar{t})}a_{R,j}^{(\bar{t})}(\theta_{R,l_{1},n})\bar{\mathbf{q}}_{l_{1},j,n}^{(r,\bar{t})}
  \end{matrix}
  \right],\label{formuA17}
\end{align}
where $\mathbf{p}_{l_{2},j,n}^{(r,\bar{t})}\in\mathbb{R}^{2\times1}$, $\bar{\mathbf{p}}_{l_{1},j,n}^{(\tilde{t},\bar{t})}\in\mathbb{R}^{2\times1}$, $o_{l_{2},j,n}^{(r,\bar{t})}$ ,$\bar{o}_{l_{1},j,n}^{(\tilde{t},\bar{t})}$,$\mathbf{q}_{l_{2},j,n}^{(r,\bar{t})}\in\mathbb{R}^{(2N(L_{2}+1)+3)\times 1}$ and $\bar{q}_{l_{1},j,n}^{(r,\bar{t})}\in\mathbb{R}^{(2N(L_{1}+1)+3)\times 1}$ are written as
\begin{align}
\mathbf{p}_{l_{2},j,n}^{(r,\bar{t})}=\left\{
 \begin{array}{lr}
 \bar{j}2\pi\frac{j}{N_{S}T_{S}}\frac{\mathbf{x}-\mathbf{b}}{\|\mathbf{x}-\mathbf{b}\|^{2}}\\
 -\bar{j}\frac{d_{R}\pi}{\lambda_{j}}(\bar{t}-1)\frac{\cos(\arccos(\frac{(\mathbf{x}-\mathbf{b})^{H}\mathbf{e}_{X}}{\|\mathbf{x}-\mathbf{b}\|^{2}})-\varphi)}{\sqrt{1-(\frac{(\mathbf{x}-\mathbf{b})^{H}\mathbf{e}_{X}}{\|\mathbf{x}-\mathbf{b}\|^{2}})^{2}}}\frac{\|\mathbf{x}-\mathbf{b}\|^{2}\mathbf{e}_{Y}-(\mathbf{x}-\mathbf{b})(\mathbf{x}-\mathbf{b})^{H}\mathbf{e}_{Y}}{\|\mathbf{x}-\mathbf{b}\|^{2}}&  \\
 +\bar{j}\frac{d_{U}\pi}{\lambda_{j}}(r-1)\frac{\cos(\arccos(\frac{(\mathbf{x}-\mathbf{b})^{H}\mathbf{e}_{X}}{\|\mathbf{x}-\mathbf{b}\|^{2}})-\varphi)}{\sqrt{1-(\frac{(\mathbf{x}-\mathbf{b})^{H}\mathbf{e}_{X}}{\|\mathbf{x}-\mathbf{b}\|^{2}})^{2}}}\frac{\|\mathbf{x}-\mathbf{b}\|^{2}\mathbf{e}_{X}-(\mathbf{x}-\mathbf{b})(\mathbf{x}-\mathbf{b})^{H}\mathbf{e}_{X}}{\|\mathbf{x}-\mathbf{b}\|^{2}}, & l_{2}=0.\\
\bar{j}2\pi\frac{j}{N_{S}T_{S}}\frac{\mathbf{u}_{l_{2},n}-\mathbf{b}}{\|\mathbf{u}_{l_{2},n}-\mathbf{b}\|^{2}}\\
+\bar{j}\frac{d_{U}\pi}{\lambda_{j}}(r-1)\frac{\cos(\arccos(\frac{(\mathbf{u}_{l_{2},n}-\mathbf{b})^{H}\mathbf{e}_{X}}{\|\mathbf{u}_{l_{2},n}-\mathbf{b}\|^{2}})-\varphi)}{\sqrt{1-(\frac{(\mathbf{u}_{l_{2},n}-\mathbf{b})^{H}\mathbf{e}_{X}}{\|\mathbf{u}_{l_{2},n}-\mathbf{b}\|^{2}})^{2}}}\frac{\|\mathbf{u}_{l_{2},n}-\mathbf{b}\|^{2}\mathbf{e}_{X}-(\mathbf{u}_{l_{2},n}-\mathbf{b})(\mathbf{u}_{l_{2},n}-\mathbf{b})^{H}\mathbf{e}_{X}}{\|\mathbf{u}_{l_{2},n}-\mathbf{b}\|^{2}}, & l_{2}>0.  
\end{array}
\right.\nonumber\\
\bar{\mathbf{p}}_{l_{1},j,n}^{(\tilde{t},\bar{t})}=\left\{
 \begin{array}{lr}
 \bar{j}2\pi\frac{j}{N_{S}T_{S}}\frac{\mathbf{b}-\mathbf{a}_{n}}{\|\mathbf{b}-\mathbf{a}_{n}\|^{2}}\\
 -\bar{j}\frac{d_{B}\pi}{\lambda_{j}}(\tilde-1)\frac{\cos(\arccos(\frac{(\mathbf{b}-\mathbf{a}_{n})^{H}\mathbf{e}_{X}}{\|\mathbf{b}-\mathbf{a}_{n}\|^{2}})-\phi_{n})}{\sqrt{1-(\frac{(\mathbf{b}-\mathbf{a}_{n})^{H}\mathbf{e}_{X}}{\|\mathbf{b}-\mathbf{a}_{n}\|^{2}})^{2}}}\frac{\|\mathbf{b}-\mathbf{a}_{n}\|^{2}\mathbf{e}_{Y}-(\mathbf{b}-\mathbf{a}_{n})(\mathbf{b}-\mathbf{a}_{n})^{H}\mathbf{e}_{Y}}{\|\mathbf{b}-\mathbf{a}_{n}\|^{2}}&  \\
 +\bar{j}\frac{d_{R}\pi}{\lambda_{j}}(\bar{t}-1)\frac{\cos(\arccos(\frac{(\mathbf{b}-\mathbf{a}_{n})^{H}\mathbf{e}_{X}}{\|\mathbf{b}-\mathbf{a}_{n}\|^{2}})-\phi_{n})}{\sqrt{1-(\frac{(\mathbf{b}-\mathbf{a}_{n})^{H}\mathbf{e}_{X}}{\|\mathbf{b}-\mathbf{a}_{n}\|^{2}})^{2}}}\frac{\|\mathbf{b}-\mathbf{a}_{n}\|^{2}\mathbf{e}_{X}-(\mathbf{b}-\mathbf{a}_{n})(\mathbf{b}-\mathbf{a}_{n})^{H}\mathbf{e}_{X}}{\|\mathbf{b}-\mathbf{a}_{n}\|^{2}}, & l_{1}=0.\\
\bar{j}2\pi\frac{n}{N_{S}T_{S}}\frac{\mathbf{r}_{l_{1},n}-\mathbf{a}_{n}}{\|\mathbf{r}_{l_{1},n}-\mathbf{a}_{n}\|^{2}}\\
+\bar{j}\frac{d_{R}\pi}{\lambda_{j}}(\bar{t}-1)\frac{\cos(\arccos(\frac{(\mathbf{r}_{l_{1},n}-\mathbf{a}_{n})^{H}\mathbf{e}_{X}}{\|\mathbf{r}_{l_{1},n}-\mathbf{a}_{n}\|^{2}})-\phi_{n})}{\sqrt{1-(\frac{(\mathbf{r}_{l_{1},n}-\mathbf{a}_{n})^{H}\mathbf{e}_{X}}{\|\mathbf{r}_{l_{1},n}-\mathbf{a}_{n}\|^{2}})^{2}}}\frac{\|\mathbf{r}_{l_{1},n}-\mathbf{a}_{n}\|^{2}\mathbf{e}_{X}-(\mathbf{r}_{l_{1},n}-\mathbf{a}_{n})(\mathbf{r}_{l_{1},n}-\mathbf{a}_{n})^{H}\mathbf{e}_{X}}{\|\mathbf{r}_{l_{1},n}-\mathbf{a}_{n}\|^{2}}, & l_{1}>0.
\end{array}
\right.\nonumber
\end{align}
\begin{align}
o_{l_{2},j,n}^{(r,\bar{t})}=\left\{
 \begin{array}{lr}
\bar{j}\frac{d_{U}\pi}{\lambda_{j}}(r-1)\cos(\arccos(\frac{(\mathbf{x}-\mathbf{b})^{H}\mathbf{e}_{X}}{\|\mathbf{x}-\mathbf{b}\|^{2}})-\omega), & l_{2}=0.\\
\bar{j}\frac{d_{U}\pi}{\lambda_{j}}(r-1)\cos(\arccos(\frac{(\mathbf{u}_{l_{2},n}-\mathbf{b})^{H}\mathbf{e}_{X}}{\|\mathbf{u}_{l_{2},n}-\mathbf{b}\|^{2}})-\varphi), & l_{2}>0.  
\end{array}
\right.\nonumber\\
\bar{o}_{l_{1},j,n}^{(\tilde{t},\bar{t})}=\left\{
 \begin{array}{lr}
\bar{j}\frac{d_{R}\pi}{\lambda_{j}}(\bar{t}-1)\cos(\arccos(\frac{(\mathbf{b}-\mathbf{a}_{n})^{H}\mathbf{e}_{X}}{\|\mathbf{b}-\mathbf{a}_{n}\|^{2}})-\phi_{n}), & l_{1}=0.\\
\bar{j}\frac{d_{R}\pi}{\lambda_{j}}(\bar{t}-1)\cos(\arccos(\frac{(\mathbf{r}_{l_{1},n}-\mathbf{a}_{n})^{H}\mathbf{e}_{X}}{\|\mathbf{r}_{l_{1},n}-\mathbf{a}_{n}\|^{2}})-\phi_{n}), & l_{1}>0. 
\end{array}
\right.\nonumber
\end{align}

\begin{align}
[\mathbf{q}_{l_{2},j,n}^{(r,\bar{t})}]_{2l-1:2l}=\left\{
 \begin{array}{lr}
\tilde{\mathbf{q}}_{l_{2},j,n}^{(r,\bar{t})}, & l=\tilde{r}N+l_{2}, \tilde{r}\in\mathbb{Z}^{+}.\\
0, & l\neq \tilde{r}N+l_{2}, \tilde{r}\in\mathbb{Z}^{+}.  
\end{array}
\right.
[\bar{\mathbf{q}}_{l_{1},j,n}^{(\tilde{t},\bar{t})}]_{2\bar{l}-1:2\bar{l}}=\left\{
 \begin{array}{lr}
\hat{\mathbf{q}}_{l_{1},j,n}^{(\tilde{t},\bar{t})}, & \bar{l}=\hat{r}N+l_{1}, \hat{r}\in\mathbb{Z}^{+}.\\
0, & \bar{l}\neq \hat{r}N+l_{1}, \hat{r}\in\mathbb{Z}^{+},
\end{array}
\right.\label{formuB21}
\end{align}
where $\tilde{\mathbf{q}}_{l_{2},j,n}^{(r,\bar{t})}$ and $\hat{\mathbf{q}}_{l_{1},j,n}^{(\tilde{t},\bar{t})}$ are expressed as
\begin{align}
&\tilde{\mathbf{q}}_{l_{2},j,n}^{(r,\bar{t})}=-\bar{j}2\pi\frac{j}{N_{S}T_{S}}\frac{\mathbf{x}-\mathbf{u}_{l_{2},n}}{\|\mathbf{x}-\mathbf{u}_{l_{2},n}\|^{2}}-\bar{j}2\pi\frac{j}{N_{S}T_{S}}\frac{\mathbf{b}-\mathbf{u}_{l_{2},n}}{\|\mathbf{b}-\mathbf{u}_{l_{2},n}\|^{2}}\nonumber\\
&-\bar{j}\frac{d_{R}\pi}{\lambda_{j}}(\bar{t}-1)\frac{\cos(\arccos(\frac{(\mathbf{b}-\mathbf{u}_{l_{2},n})^{H}\mathbf{e}_{X}}{\|\mathbf{b}-\mathbf{u}_{l_{2},n}\|^{2}})-\varphi)}{\sqrt{1-(\frac{(\mathbf{b}-\mathbf{u}_{l_{2},n})^{H}\mathbf{e}_{X}}{\|\mathbf{b}-\mathbf{u}_{l_{2},n}\|^{2}})^{2}}}\frac{\|\mathbf{b}-\mathbf{u}_{l_{2},n}\|^{2}\mathbf{e}_{Y}-(\mathbf{b}-\mathbf{u}_{l_{2},n})(\mathbf{b}-\mathbf{u}_{l_{2},n})^{H}\mathbf{e}_{Y}}{\|\mathbf{b}-\mathbf{u}_{l_{2},n}\|^{2}}\nonumber\\
&+\bar{j}\frac{d_{U}\pi}{\lambda_{n}}(r-1)\frac{\cos(\arccos(\frac{(\mathbf{x}-\mathbf{u}_{l_{2},n})^{H}\mathbf{e}_{X}}{\|\mathbf{x}-\mathbf{u}_{l_{2},n}\|^{2}})-\omega)}{\sqrt{1-(\frac{(\mathbf{x}-\mathbf{u}_{l_{2},n})^{H}\mathbf{e}_{X}}{\|\mathbf{x}-\mathbf{u}_{l_{2},n}\|^{2}})^{2}}}\frac{\|\mathbf{x}-\mathbf{u}_{l_{2},n}\|^{2}\mathbf{e}_{X}-(\mathbf{x}-\mathbf{u}_{l_{2},n})(\mathbf{x}-\mathbf{u}_{l_{2},n})^{H}\mathbf{e}_{X}}{\|\mathbf{x}-\mathbf{u}_{l_{2},n}\|^{2}}\nonumber\\
&\hat{\mathbf{q}}_{l_{1},j,n}^{(\tilde{t},\bar{t})}=-\bar{j}2\pi\frac{j}{N_{S}T_{S}}\frac{\mathbf{b}-\mathbf{r}_{l_{1},n}}{\|\mathbf{b}-\mathbf{r}_{l_{1},n}\|^{2}}-\bar{j}2\pi\frac{j}{N_{S}T_{S}}\frac{\mathbf{r}_{l_{1},n}-\mathbf{a}_{n}}{\|\mathbf{r}_{l_{1},n}-\mathbf{a}_{n}\|^{2}}\nonumber\\
&-\bar{j}\frac{d_{B}\pi}{\lambda_{j}}(\tilde{t}-1)\frac{\cos(\arccos(\frac{(\mathbf{b}-\mathbf{r}_{l_{1},n})^{H}\mathbf{e}_{X}}{\|\mathbf{b}-\mathbf{r}_{l_{1},n}\|^{2}})-\varphi)}{\sqrt{1-(\frac{(\mathbf{b}-\mathbf{r}_{l_{1},n})^{H}\mathbf{e}_{X}}{\|\mathbf{b}-\mathbf{r}_{l_{1},n}\|^{2}})^{2}}}\frac{\|\mathbf{r}_{l_{1},n}-\mathbf{a}_{n}\|^{2}\mathbf{e}_{Y}-(\mathbf{r}_{l_{1},n}-\mathbf{a}_{n})(\mathbf{r}_{l_{1},n}-\mathbf{a}_{n})^{H}\mathbf{e}_{Y}}{\|\mathbf{r}_{l_{1},n}-\mathbf{a}_{n}\|^{2}}\nonumber\\
&+\bar{j}\frac{d_{R}\pi}{\lambda_{j}}(\bar{t}-1)\frac{\cos(\arccos(\frac{(\mathbf{b}-\mathbf{r}_{l_{1},n})^{H}\mathbf{e}_{X}}{\|\mathbf{b}-\mathbf{r}_{l_{1},n}\|^{2}})-\varphi)}{\sqrt{1-(\frac{(\mathbf{b}-\mathbf{r}_{l_{1},n})^{H}\mathbf{e}_{X}}{\|\mathbf{b}-\mathbf{r}_{l_{1},n}\|^{2}})^{2}}}\frac{\|\mathbf{b}-\mathbf{r}_{l_{1},n}\|^{2}\mathbf{e}_{X}-(\mathbf{b}-\mathbf{r}_{l_{1},n})(\mathbf{b}-\mathbf{r}_{l_{1},n})^{H}\mathbf{e}_{X}}{\|\mathbf{b}-\mathbf{r}_{l_{1},n}\|^{2}}.\label{formu89}
\end{align}
Thus, $\mathrm{MSE}(\mathbf{w},\boldsymbol{\theta},\mathbf{g}(t))=\mathbb{E}_{\mathbf{h}[t],\mathbf{n}}\{\|\hat{\boldsymbol{\rho}}-\boldsymbol{\rho}\|^{2}\}\geq \mathrm{trace}(\boldsymbol{\mathcal{J}}(\boldsymbol{\kappa},\mathbf{h}[t];\mathbf{w},\boldsymbol{\theta},\mathbf{g}[t])^{-1})$ and $\mathrm{MSE}(\mathbf{w},\boldsymbol{\theta},\mathbf{h}[t])=\mathbb{E}_{\mathbf{g}[t],\mathbf{n}}\{\|\hat{\boldsymbol{\rho}}-\boldsymbol{\rho}\|^{2}\}\geq\mathrm{trace}(\boldsymbol{\mathcal{I}}(\boldsymbol{\kappa},\mathbf{g}[t];\mathbf{w},\boldsymbol{\theta},\mathbf{h}[t])^{-1})$, the $\textbf{Theorem~1}$ is proved. 
\section{Proof of \textbf{Theorem~\ref{The32}}}\label{appB1}
According to \cite{kay1993fundamentals}, we have
\begin{align}
\|\mathbf{\Gamma}((\boldsymbol{\kappa}_{2}+\bar{\boldsymbol{\kappa}}_{2}),\boldsymbol{\theta},\mathbf{w},\mathbf{g}[t])-\mathbf{\Gamma}(\boldsymbol{\kappa}_{2},\boldsymbol{\theta},\mathbf{w},\mathbf{g}[t])\|^{2}=\|\nabla^{H}_{\boldsymbol{\kappa}_{2}}\mathbf{\Gamma}(\boldsymbol{\kappa}_{2},\boldsymbol{\theta},\mathbf{w},\mathbf{g}[t])\bar{\boldsymbol{\kappa}}_{2}+\mathcal{O}(\|\boldsymbol{\kappa}_{2}\|^{2})\|^{2}.\label{formu90}
\end{align}
Based on triangle inequality, the right hand side of (\ref{formu90}) is simplified as
\begin{align}
\|\nabla^{H}_{\boldsymbol{\kappa}_{2}}\mathbf{\Gamma}(\boldsymbol{\kappa}_{2},\boldsymbol{\theta},\mathbf{w},\mathbf{g}[t])\bar{\boldsymbol{\kappa}}_{2}+\mathcal{O}(\|\boldsymbol{\kappa}_{2}\|^{2})\|^{2}\geq \|\nabla^{H}_{\boldsymbol{\kappa}_{2}}\mathbf{\Gamma}(\boldsymbol{\kappa}_{2},\boldsymbol{\theta},\mathbf{w},\mathbf{g}[t])\bar{\boldsymbol{\kappa}}_{2}\|^{2}-\mathcal{O}(\|\boldsymbol{\kappa}_{2}\|^{2}).\label{formu91}
\end{align}
Thus, we have
\begin{align}
\|\nabla^{H}_{\boldsymbol{\kappa}_{2}}\mathbf{\Gamma}(\boldsymbol{\kappa}_{2},\boldsymbol{\theta},\mathbf{w},\mathbf{g}[t])\bar{\boldsymbol{\kappa}}_{2}+\mathcal{O}(\|\boldsymbol{\kappa}_{2}\|^{2})\|^{2}+\mathcal{O}(\|\boldsymbol{\kappa}_{2}\|^{2})\geq\|\nabla^{H}_{\boldsymbol{\kappa}_{2}}\mathbf{\Gamma}(\boldsymbol{\kappa}_{2},\boldsymbol{\theta},\mathbf{w},\mathbf{g}[t])\bar{\boldsymbol{\kappa}}_{2}\|^{2}.\label{formu92}
\end{align}
According to Cauchy-Schwarz inequality, the following inequality is expressed as
\begin{align}
\|\nabla^{H}_{\boldsymbol{\kappa}_{2}}\mathbf{\Gamma}(\boldsymbol{\kappa}_{2},\boldsymbol{\theta},\mathbf{w},\mathbf{g}[t])\bar{\boldsymbol{\kappa}}_{2}\|^{2}\geq C_{1} \|\nabla^{H}_{\boldsymbol{\kappa}_{2}}\mathbf{\Gamma}(\boldsymbol{\kappa}_{2},\boldsymbol{\theta},\mathbf{w},\mathbf{g}[t])\|^{2}\|\bar{\boldsymbol{\kappa}}_{2}\|^{2},~~C_{1}\geq 0.\label{formu93}
\end{align}
Therefore, we have
\begin{align}
\|\nabla^{H}_{\boldsymbol{\kappa}_{2}}\mathbf{\Gamma}(\boldsymbol{\kappa}_{2},\boldsymbol{\theta},\mathbf{w},\mathbf{g}[t])\bar{\boldsymbol{\kappa}}_{2}+\mathcal{O}(\|\boldsymbol{\kappa}_{2}\|^{2})\|^{2}+\mathcal{O}(\|\boldsymbol{\kappa}_{2}\|^{2})\geq C_{1} \|\nabla^{H}_{\boldsymbol{\kappa}_{2}}\mathbf{\Gamma}(\boldsymbol{\kappa}_{2},\boldsymbol{\theta},\mathbf{w},\mathbf{g}[t])\|^{2}\|\bar{\boldsymbol{\kappa}}_{2}\|^{2}.\label{formu94}
\end{align}
Then, we get
\begin{align}
\|\bar{\boldsymbol{\kappa}}_{2}\|^{2}\leq \frac{\|\mathbf{\Gamma}((\boldsymbol{\kappa}_{2}+\bar{\boldsymbol{\kappa}}_{2}),\boldsymbol{\theta},\mathbf{w},\mathbf{g}[t])-\mathbf{\Gamma}(\boldsymbol{\kappa}_{2},\boldsymbol{\theta},\mathbf{w},\mathbf{g}[t])\|^{2}+\mathcal{O}(\|\boldsymbol{\kappa}_{2}\|^{2})}{C_{1} \|\nabla^{H}_{\boldsymbol{\kappa}_{2}}\mathbf{\Gamma}(\boldsymbol{\kappa}_{2},\boldsymbol{\theta},\mathbf{w},\mathbf{g}[t])\|^{2}}.\label{formu95}
\end{align}
When the estimation error is $0$, (\ref{formu95}) is rewritten as
\begin{align}
\|\bar{\boldsymbol{\kappa}}_{2}\|^{2}\leq\frac{\mathcal{O}(\|\boldsymbol{\kappa}_{2}\|^{2})}{C_{1} \|\nabla^{H}_{\boldsymbol{\kappa}_{2}}\mathbf{\Gamma}(\boldsymbol{\kappa}_{2},\boldsymbol{\theta},\mathbf{w},\mathbf{g}[t])\|^{2}}.\label{formu96}
\end{align}
When the estimation error is not $0$, since $\boldsymbol{\kappa}_{2}$ is the optimal solution and $\mathbf{\Gamma}(\boldsymbol{\kappa}_{2},\boldsymbol{\theta},\mathbf{w},\mathbf{g}[t])$ is noiseless, we have
\begin{align}
\|\mathbf{\Gamma}((\boldsymbol{\kappa}_{2}+\bar{\boldsymbol{\kappa}}_{2}),\boldsymbol{\theta},\mathbf{w},\mathbf{g}[t])-\mathbf{\Gamma}(\boldsymbol{\kappa}_{2},\boldsymbol{\theta},\mathbf{w},\mathbf{g}[t])\|^{2}\leq\|\mathbf{n}[k]\|^{2}.\label{formu97}
\end{align}
Thus, we have
\begin{align}
\|\bar{\boldsymbol{\kappa}}_{2}\|^{2}\leq\frac{\|\mathbf{n}[k]\|^{2}+\mathcal{O}(\|\boldsymbol{\kappa}_{2}\|^{2})}{C_{1} \|\nabla^{H}_{\boldsymbol{\kappa}_{2}}\mathbf{\Gamma}(\boldsymbol{\kappa}_{2},\boldsymbol{\theta},\mathbf{w},\mathbf{g}[t])\|^{2}}.\label{formu98}
\end{align}
According to (\ref{formu96}) and (\ref{formu98}), we have
\begin{align}
\|\bar{\boldsymbol{\kappa}}_{2}\|^{2}\sim\mathcal{O}\left(\frac{\|\mathbf{n}[k]\|^{2}}{ \|\nabla^{H}_{\boldsymbol{\kappa}_{2}}\mathbf{\Gamma}(\boldsymbol{\kappa}_{2},\boldsymbol{\theta},\mathbf{w},\mathbf{g}[t])\|^{2}}\right).
\end{align}
Similarly, the estimation error of (\ref{formu18}) satisfy the following equality  
\begin{align}
\|\bar{\boldsymbol{\kappa}}_{1}\|^{2}\leq\frac{\|\mathbf{n}[k]\|^{2}}{\|\nabla^{H}_{\boldsymbol{\kappa}_{1}}\mathbf{\Lambda}(\boldsymbol{\kappa}_{1},\boldsymbol{\theta},\mathbf{w},\mathbf{h}[t])\|^{2}}.\label{formu99}
\end{align}
\section{Proof of \textbf{Theorem~\ref{The33}}}\label{appB2}
According to (\ref{formu46}),  $\boldsymbol{\mathcal{P}}$ is expressed as
\begin{align}
\boldsymbol{\mathcal{P}}=
\mathbf{G}_{n}^{T}[t,j]\boldsymbol{\Theta}^{T}\left(\sum_{l_{1},r}\frac{\eta_{l}^{2}\mathbf{U}_{l,j,n}^{(r)H}\mathbf{B}^{H}(\mathbf{w}^{(i)})\mathbf{B}(\mathbf{w}^{(i)})\mathbf{U}_{l,j,n}^{(r)}(\boldsymbol{\kappa})}{\sigma^{2}}\right)\boldsymbol{\Theta}^{*}\mathbf{G}_{n}^{*}[t,j]-\nonumber\\
\mathbf{G}_{n}^{T}[t,j]\boldsymbol{\Theta}^{T}\left(\sum_{l_{1},r}\frac{\mathbf{V}_{j,n}^{(r)H}\mathbf{E}^{H}(\mathbf{w}^{(i)})\mathbf{E}(\mathbf{w}^{(i)})\mathbf{V}_{j,n}^{(r)}(\boldsymbol{\kappa})}{\sigma^{2}}\right)\boldsymbol{\Theta}^{*}\mathbf{G}_{n}^{*}[t,j]-\nonumber\\
\left(\sum_{l_{2},r}\frac{\eta_{l}^{2}\bar{\mathbf{U}}_{l,j,n}^{(r)H}\bar{\mathbf{B}}^{H}(\mathbf{w}^{(i)})\bar{\mathbf{B}}(\mathbf{w}^{(i)})\bar{\mathbf{U}}_{l,j,n}^{(r)}(\boldsymbol{\kappa})}{\sigma^{2}}\right)-\left(\sum_{l_{2},r}\frac{\bar{\mathbf{V}}_{j,n}^{(r)H}\bar{\mathbf{E}}^{H}(\mathbf{w}^{(i)})\bar{\mathbf{E}}(\mathbf{w}^{(i)})\bar{\mathbf{V}}_{j,n}^{(r)}(\boldsymbol{\kappa})}{\sigma^{2}}\right).\label{formuA_46}
\end{align}
For matrix $\boldsymbol{\mathcal{P}}$, we assume $\bar{\lambda}_{max}$ is the largest eigenvalue of $\boldsymbol{\mathcal{P}}$, and $\bar{\mathbf{w}}_{n}[j,m]^{(i+1)}$ is the principal eigenvector, therefore, we have
\begin{align}
(\bar{\mathbf{w}}_{n}[j,m]^{(i+1)})^{H}\boldsymbol{\mathcal{P}}\bar{\mathbf{w}}_{n}[j,m]^{(i+1)}=\bar{\lambda}_{max}(\bar{\mathbf{w}}_{n}[j,m]^{(i+1)})^{H}\bar{\mathbf{w}}_{n}[j,m]^{(i+1)}.\label{formu101}
\end{align}
According to (\ref{formu46}) and (\ref{formu101}), we have
\begin{align}
(\mathbf{d}^{H}_{n}[j,m])^{(i+1)}\nabla_{\mathbf{w}_{n}^{(i)}[j,m]}\bar{f}_{S}(\mathbf{w}_{n}^{(i)}[j,m])=\nonumber\\
-2(\bar{\mathbf{w}}^{H}_{n}[j,m])^{H}\frac{(\|\mathbf{w}_{n}^{(i)}[j,m]\|^{2}\mathbf{I}_{N_{T}}-\mathbf{w}_{n}^{*(i)}[j,m]\mathbf{w}_{n}^{T(i)}[j,m])}{\|\mathbf{w}_{n}^{(i)}[j,m]\|^{4}}\mathbf{B}(\mathbf{w})\mathbf{w}_{n}^{*(i)}[j,m]\nonumber\\
=\frac{(\bar{\mathbf{w}}^{H}_{n}[j,m])^{H}\mathbf{w}_{n}^{*(i)}[j,m]}{\|\mathbf{w}_{n}^{(i)}[j,m]\|^{2}}(\frac{\mathbf{w}_{n}^{*(i)}[j,m]^{T}\mathbf{B}(\mathbf{w})\mathbf{w}_{n}^{*(i)}[j,m]}{\|\mathbf{w}_{n}^{(i)}[j,m]\|^{2}}-\bar{\lambda}_{max})\leq 0\label{formu102}
\end{align}
where the largest value of expression $\frac{\mathbf{w}_{n}^{*(i)}[j,m]^{T}\mathbf{B}(\mathbf{w})\mathbf{w}_{n}^{*(i)}[j,m]}{\|\mathbf{w}_{n}^{(i)}[j,m]\|^{2}}$ is $\bar{\lambda}_{max}$. (\ref{formu102}) shows that without considering the internal constraint structure of the objective function or making the objective function reach the maximum, the inner product of gradient vector and direction vector is always less than 0. Thus, according to \cite{shi2004convergence, zhang2006global}, when the inner product of gradient vector and direction vector is always less than 0, the gradient descent optimization algorithm based on Armijo rule is monotonous and converges to a stable point. Proof of $\textbf{Theorem}~3$ has been finished.


\end{document}